\documentclass[aps,pre,preprint]{revtex4}
\usepackage{graphicx}
\begin{document}

\title{Nanoscale simulations of directional locking}

\author{Joel Koplik}
\email[]{koplik@sci.ccny.cuny.edu}
\affiliation{Benjamin Levich Institute and Department of Physics\\
City College of the City University of New York, New York, NY 10031}

\author{German Drazer}
\email[]{drazer@jhu.edu}
\affiliation{Department of Chemical and Biomolecular Engineering\\
Johns Hopkins University, Baltimore, MD 21218}
\date{\today}

\begin{abstract}
When particles suspended in a fluid are driven through a regular lattice 
of cylindrical obstacles, the particle motion is usually not simply in the 
direction of the force, and in the high P\'eclet number limit particle
trajectories tend to lock along certain lattice directions.  By means of
molecular dynamics simulations we show that this effect persists in the 
presence of molecular diffusion for nanoparticle flows, provided the 
P\'eclet number is not too small.  We examine the effects of varying
particle and obstacle size, the method of forcing, solid roughness, 
and particle concentration.  While we observe trajectory locking in all 
cases, the {\em degree} of locking varies with particle size and these flows
may have application as a separation technique.
\end{abstract}

\pacs{}

\maketitle

\section{Introduction} \label{intro}

Macroscopic particles suspended in a fluid and driven through a
periodic lattice of solid obstacles tend to move in (commensurate)
periodic trajectories. In general, these periodic trajectories are
not aligned with the driving force.\cite{FrechetteDrazer09,BalvinFrechette09}  In fact, the average
trajectory angle typically remains locked to a given lattice direction
over finite intervals of the forcing angle ({\it directional
locking}). The locking direction $\alpha$ ($=\arctan(q/p)$, where
$(p,q)$ is a lattice direction) and its relation to the forcing
angle $\theta$ depends on the properties of the particles. Therefore,
particles of different species driven by the same external force
could migrate at different angles with respect to the underlying
lattice of solid obstacles. This behavior raises the possibility of 
using periodic obstacle lattices 
to provide deterministic and continuous separation of
suspended particles. In fact, we have experimentally shown the
possibility to separate sedimenting particles by size as they move
through an ordered array of cylindrical obstacles created with
LEGO$^{\tiny\textregistered}$ pegs on a LEGO$^{\tiny\textregistered}$
board.\cite{BalvinFrechette09}

Experiments on 
colloidal particles moving through periodic arrays have shown
analogous directional-locking behavior in the near-deterministic limit
of large P\'eclet numbers.\cite{HuangSturm04}  Moreover, a separation
technique based on the deterministic motion of suspended particles
in these periodic devices has been successfully used to fractionate
colloidal particles of different size, \cite{HuangSturm04}, isolate
blood plasma from other blood components \cite{DavisAustin06},
separate and label cells \cite{MortonAustin08} as well as to isolate
selected cell types for tissue engineering applications.\cite{GreenMurthy09}  
Similar locking behavior is observed in the 
transport of suspended colloidal particles through spatially periodic, 
smooth, two-dimensional energy landscapes, arising from an array of 
optical tweezers.\cite{KordaGrier02}  
Introducing an optical lattice
in a microfluidic channel has also been shown to selectively induce
locked-in trajectories that deflect specific particle species in a
direction different from the average flow, thus providing a means
for lateral fractionation based on the dielectric properties of the
particles.\cite{MacDonaldDholakia03,DholakiaMacDonald09}  An
analogous optical fractionation method is based on the  presence
of locked-trajectories depending on particle size.\cite{LadavacGrier04,RoichmanGrier07}

Simulations which faithfully incorporate all of the relevant experimental 
features of colloidal particles -- finite particle and obstacle size,
accurate hydrodynamics and Brownian motion -- are not as yet available.
The motion of tracer particles moving through an
array of solid obstacles was recently investigated using a Fokker-Planck
equation approach, and directional
locking was observed at large P\'eclet numbers.\cite{HerrmannDrazer09}
However, these simulations do not account for particle-solid
hydrodynamic interactions or other particle size effects. The case
of macroscopic particles was recently investigated by means of
Stokesian dynamics simulations, showing that directional locking
occurs in the presence of irreversible particle-obstacle interactions.\cite{FrechetteDrazer09}
These irreversible forces, on the other
hand, are added {\it ad hoc}, typically to prevent overlap in the
Stokesian dynamics simulations. In addition, these simulations
correspond to the limit of zero inertia and deterministic motion.
Directional locking (or phase-locking) has also been observed in 
simulations of a number of related transport phenomena, including 
driven vortex
lattices,\cite{ReichhardtNori99} the motion of overdamped particles
in periodic force fields,\cite{PeltonGrier04} particles driven
through a colloidal lattice,\cite{ReichhardtReichhardt04} particles
moving on crystalline surfaces,\cite{LacastaLindenberg05} and
periodic energy landscapes in general.\cite{LacastaLindenberg06}
Most of these simulations, however, investigate the motion in an
external field and not the interaction of suspended particles with
solid obstacles. 

In this paper we use molecular dynamics simulations to show that 
directional locking occurs in the transport of nanoparticles,
provided the P\'eclet number is not too small. 
These simulations allow us to examine the complex
interplay between Brownian motion and different driving fields,
together with hydrodynamic interactions (with possible inertia
contributions), and other finite size effects such as the atomic-scale
roughness of the particles.  In addition, molecular dynamics
simulations also allow us to investigate the effect that particle-particle
interactions have on the separation behavior at different particle
concentrations.  Moreover, at low concentrations
and for certain orientations of the driving force, we show that
particles of different size move in different directions.

\section{Methods} \label{meth}

The calculations are based on standard molecular dynamics techniques
\cite{AllenTildesley87} for atoms interacting via Lennard-Jones 
(LJ) potentials:  
\begin{equation}
V_{\rm LJ}(r) = 4\,\epsilon\, \left[ \left( {r\over\sigma} \right)^{-12} -
c_{ij}\, \left( {r\over\sigma} \right)^{-6}\ \right],
\label{eq:lj}
\end{equation}
cut off at $r_c=2.5\sigma$.  The coefficient $c_{ij}$ will be used to vary
the interaction according to the atomic species $i$ and $j$ involved.  We 
have a three component system consisting of a monatomic suspending liquid (L), 
mobile nearly-spherical particles (P) composed of atoms rigidly held together, 
and cylindrical obstacles (O) composed of atoms fixed in place.  The 
interaction coefficients are chosen to have a ``standard'' liquid,
$c_{LL}=1$, which completely wets both solids, $c_{LO}=c_{LP}=1$, and only a 
short-distance repulsion between the particles and the obstacles, $c_{OP}=0$.
We express all dimensional quantities in terms of  
the energy scale $\epsilon$, the core size $\sigma$, and the natural time 
scale $\tau\equiv \sigma(m/\epsilon)^{1/2}$, where $m$ is the common mass of 
all atoms.  Typical values are $\epsilon \sim 10^{-21}\,$J, $\sigma \sim
0.3\,$nm and $\tau \sim 2\,$ps.  

\begin{figure}
\begin{center}
\includegraphics[width=0.4\linewidth]{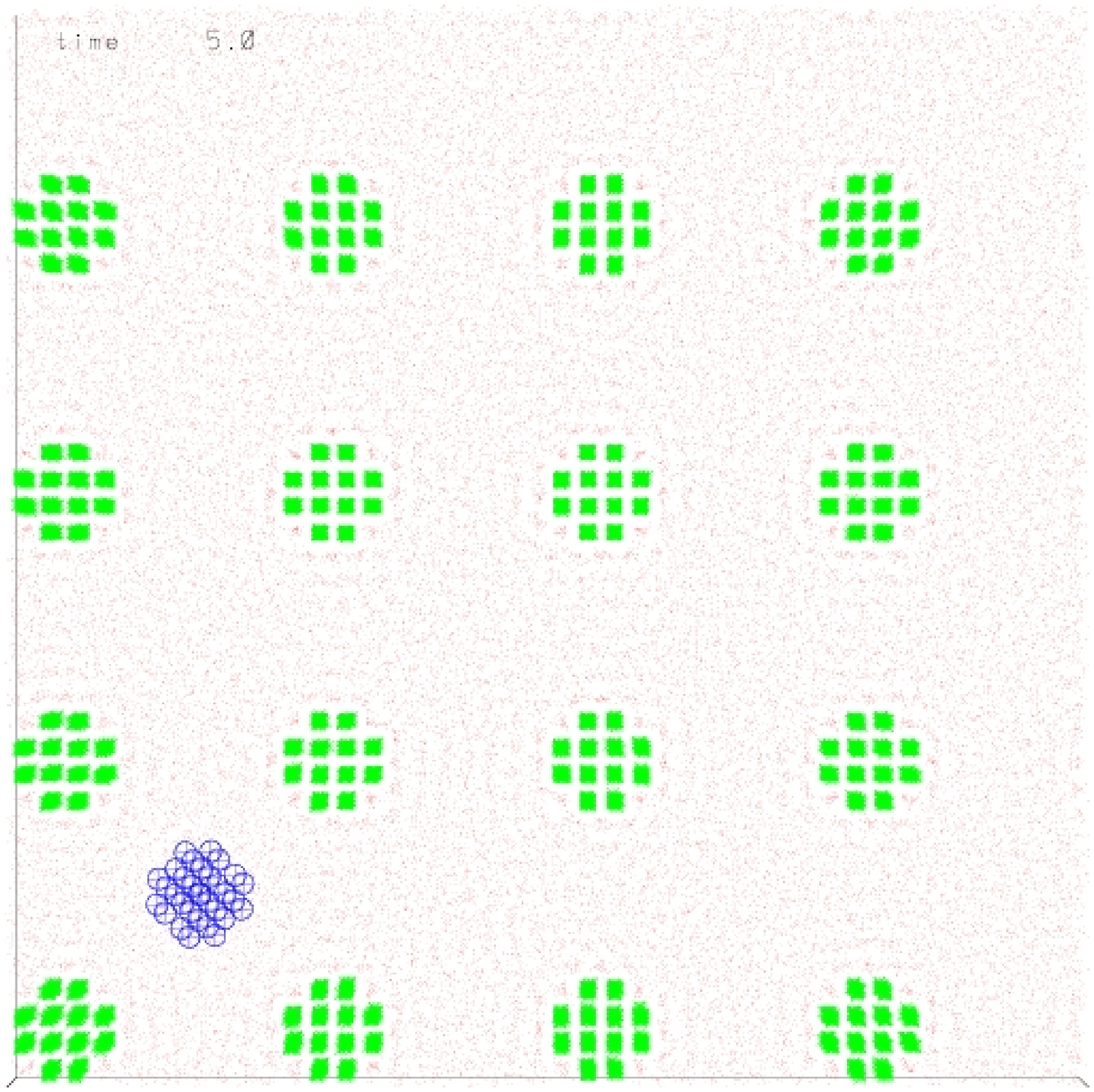}\hspace{0.2in}
\includegraphics[width=0.4\linewidth]{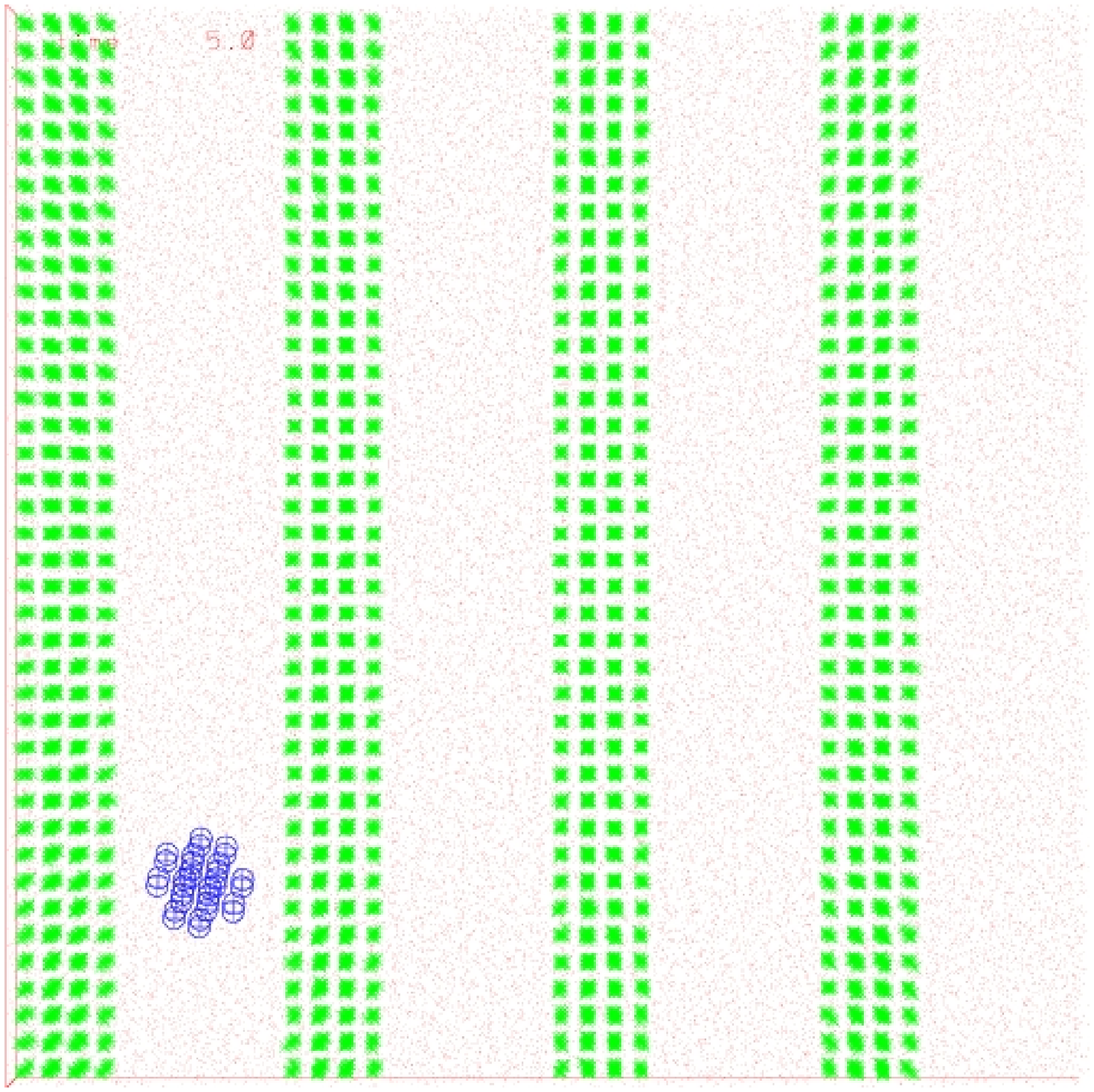}
\end{center}
\caption{\label{fig:start} (Color online)
Snapshot of the atomic positions for a simulation involving a
$4\times4$ array of cylindrical obstacles of radius 2 (green) and a single 
mobile sphere (blue), also of radius 2, suspended in a monatomic solvent
(red dots). Left: $x$-$y$ plane in which the force is applied; right: 
$x$-$z$ plane.}
\end{figure}

The initial atomic
positions form a cubic lattice of density $\rho=0.8\sigma^{-3}$. The
particle is defined by selecting all atoms within a prescribed radius of
some initial center, while the obstacles are centered on a square lattice in
the $x$-$y$ plane, and consist of all atoms whose $(x,y)$ coordinates lie
within a second prescribed radius of one of the centers.  An example of the
atomic positions after equilibration is given in Fig.~\ref{fig:start}, 
showing two
orthogonal views of a $4\times4$ lattice of cylindrical obstacles of radius 2, 
along with a particle of radius 2.  
The simulated region is a cube 43.09$\sigma$ on a side, containing 56,288 
solvent atoms, 16 obstacles with 480 atoms each, and the mobile sphere
contains 32 atoms.  
Because they are composed of sets of distinct atoms the particle and obstacle 
automatically have rough surfaces, and we discuss the effects of varying the
degree of roughness below.  Periodic boundary conditions are applied in all 
three directions;  
the spacing between the obstacles must be commensurate with the size of the
simulation box so that the obstacle lattice has the same periodicity, and 
in the case shown in the figure the ratio is 1/4.  Although 
the particle could fully sample the periodic geometry in a smaller simulation
involving just one unit cell of the obstacle lattice, the larger system
provides a larger separation between the particle and its periodic images 
and the particle's motion is correspondingly less influenced by the ``wake''
of the images.

The motion of the atoms follows the Newtonian equations appropriate to 
each species.  Each fluid atom accelerates according to the force exerted 
by {\em all} other atoms within interaction range. The particle moves
rigidly, so that its center of mass translates under the sum of the 
LJ forces applied to its individual atoms and rotates in 
response to the sum of the torques on the individual atoms about the center 
of mass.  The orientation of the particle and its rotational motion is
conveniently described by the Euler equations expressed in terms of 
quaternion variables.\cite{EvansMurad77}  The atoms
in the obstacles are simply fixed in place.  The particle is driven 
through the obstacle array by adding an external force to its LJ
interactions in a
prescribed direction in the $x$-$y$ plane, or alternatively such a force is
applied to each suspending fluid atom to produce a pressure driven
flow in the solvent in which the particle is advected.

During the simulation the temperature of the solvent liquid is maintained at
$T=1.0\epsilon/k_B$ by a Nos\'e-Hoover thermostat.  In principle, the atoms
in the particle and obstacle should also have thermal fluctuations about their
rigid-lattice positions, but the displacements in a solid are small and this
complication has been omitted in most cases.  Allowing thermal motion of the
obstacle atoms presents little computational difficulty, and can be
accomplished by using a stiff linear spring to tether these atoms to their
initial lattice sites and allowing them to move in response to the
LJ interactions with other nearby atoms.  We have added such
thermal motion to the obstacles in a few test cases and, outside of an
interesting special situation discussed below, we do not observe any
significant change in the particle motion. Adding thermal motion to the
atoms of the mobile particle is more difficult because the particle's moment 
of inertia would vary in time and complicate the solution 
of the Euler equations.

In these calculations the directed motion of the particle is modified by the
Brownian motion resulting from its interaction with the suspending fluid 
atoms, and the competition may be quantified by the P\'eclet number. 
Normally one defines $\rm{Pe}=Ua/D_m$ where $U$, $a$ and $D_m$ are the average
velocity, radius and molecular diffusivity of the particle, respectively,
but this form is awkward here because the velocity is a result of the
simulation and measuring the diffusivity in this geometry would require 
extensive additional simulations.  (The motion in $z$ is diffusive, but the
diffusion coefficient is modified from that in pure fluid by the interaction
with the rough obstacles.) However, a simpler characterization is
possible if we assume that a simple linear drag law applies,
$\bf{U}=\gamma\bf{F}$, and that the drag coefficient satisfies the 
Stokes-Einstein relation $D_m=k_BT/\gamma$, in which case $\rm{Pe}=Fa/k_BT$.
In fact, as we shall see below, $\gamma$ depends on the orientation of
$\bf{F}$ with respect to the obstacle lattice so this argument is only
approximate, and the latter form for $\rm{Pe}$ is just a convenient parametrization.

\section{Results}\label{results}

We begin with simulations of the system shown in Fig.~\ref{fig:start},
applying a force 
\begin{equation}
{\bf{F}}=(F_0\cos\theta,F_0\sin\theta,0) \label{force}
\end{equation}
to the particle.  The position of the particle's center of mass is recorded
at 1$\tau$ intervals, and the subsequent analysis is based on the $x$ and $y$
components of these trajectories.  Because of the symmetry of the square 
obstacle lattice, it suffices to consider forcing angles $\theta$ between 0 and
45$^\circ$.  

\subsection{Local behavior}
\label{ss:local}

\begin{figure}
\begin{center}
\includegraphics[width=0.45\linewidth]{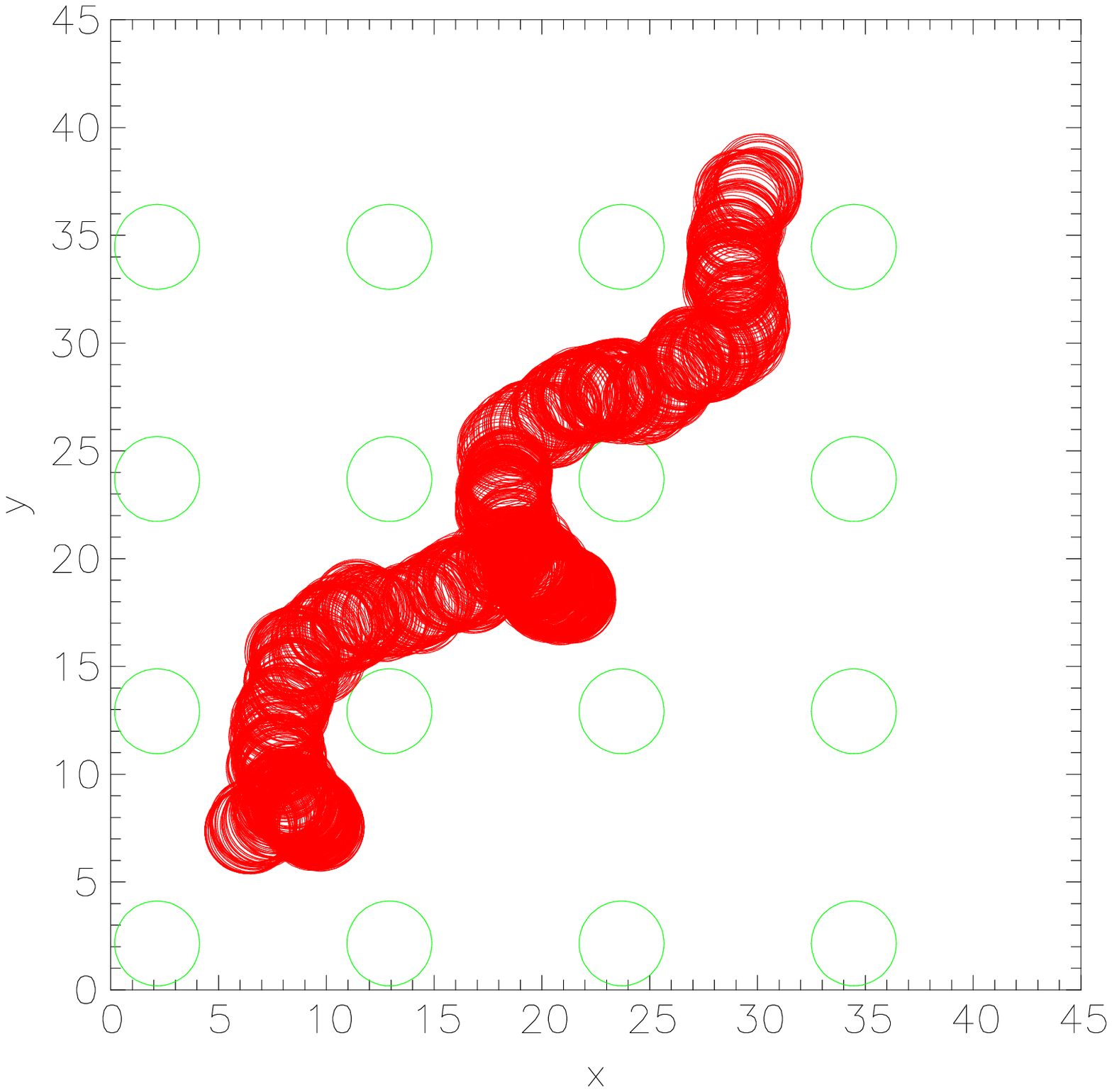}\hspace{0.2in}
\includegraphics[width=0.45\linewidth]{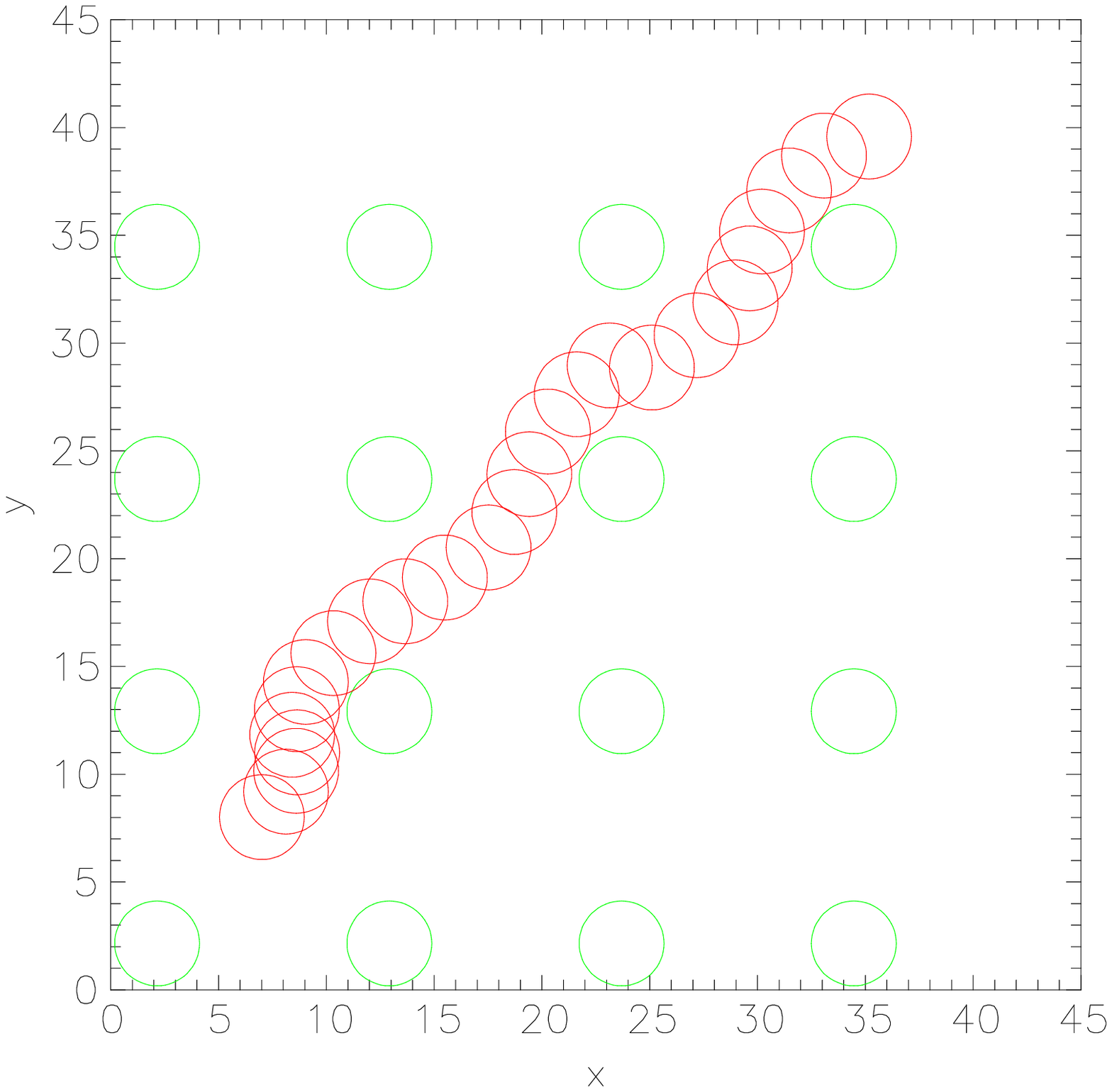}
\end{center}
\caption{\label{fig:finetraj} (Color online)
Early stages of the trajectories of a particle forced at $\theta=48^\circ$,
shown as a sequence of circles (red) moving through cylindrical obstacles
(green). Left: $F_0=5$ until 2250$\tau$; Right: $F_0=200$ until 25$\tau$.}
\end{figure}
\begin{figure}
\begin{center}
\includegraphics[width=0.45\linewidth]{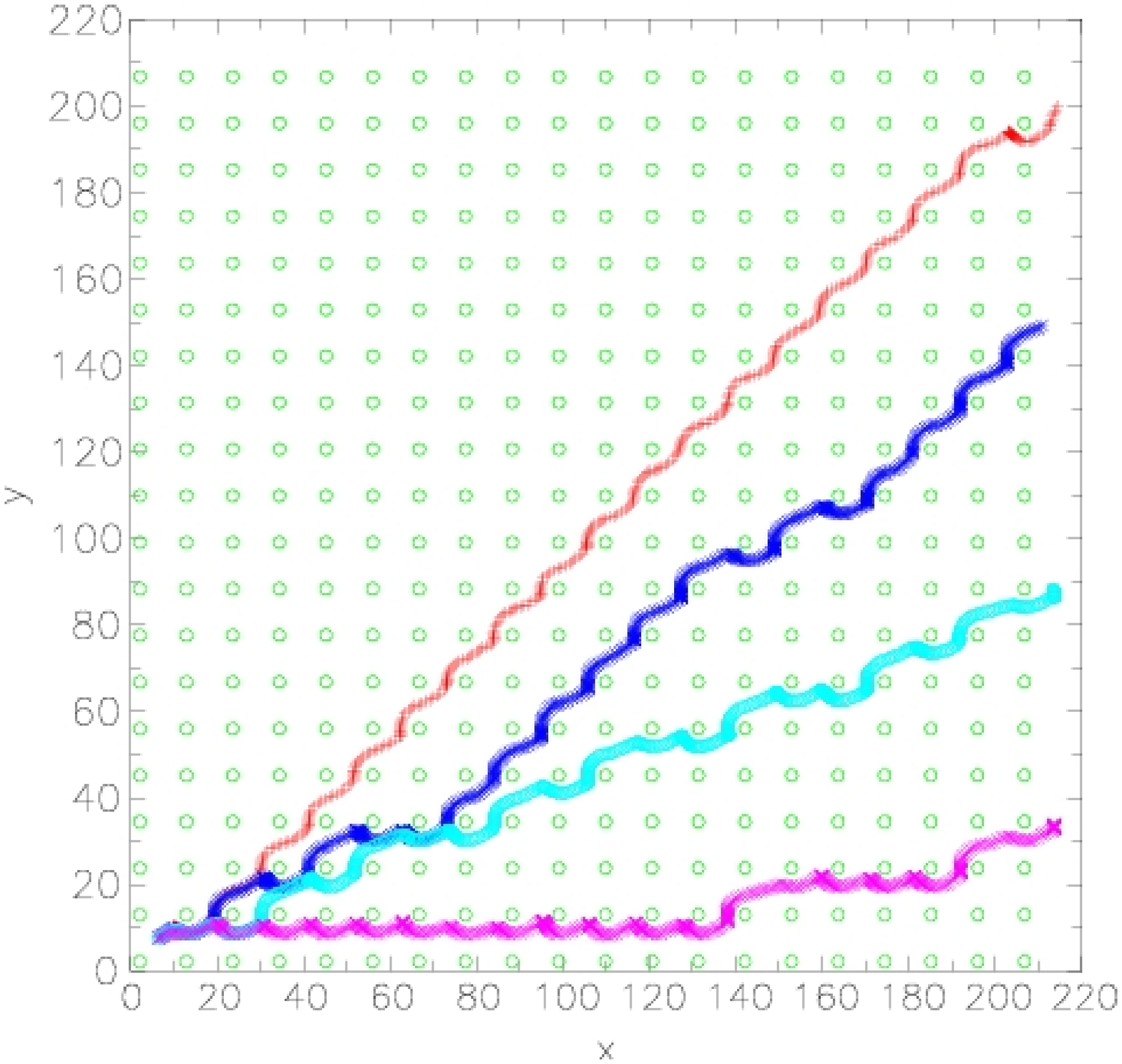}\hspace{0.2in}
\includegraphics[width=0.45\linewidth]{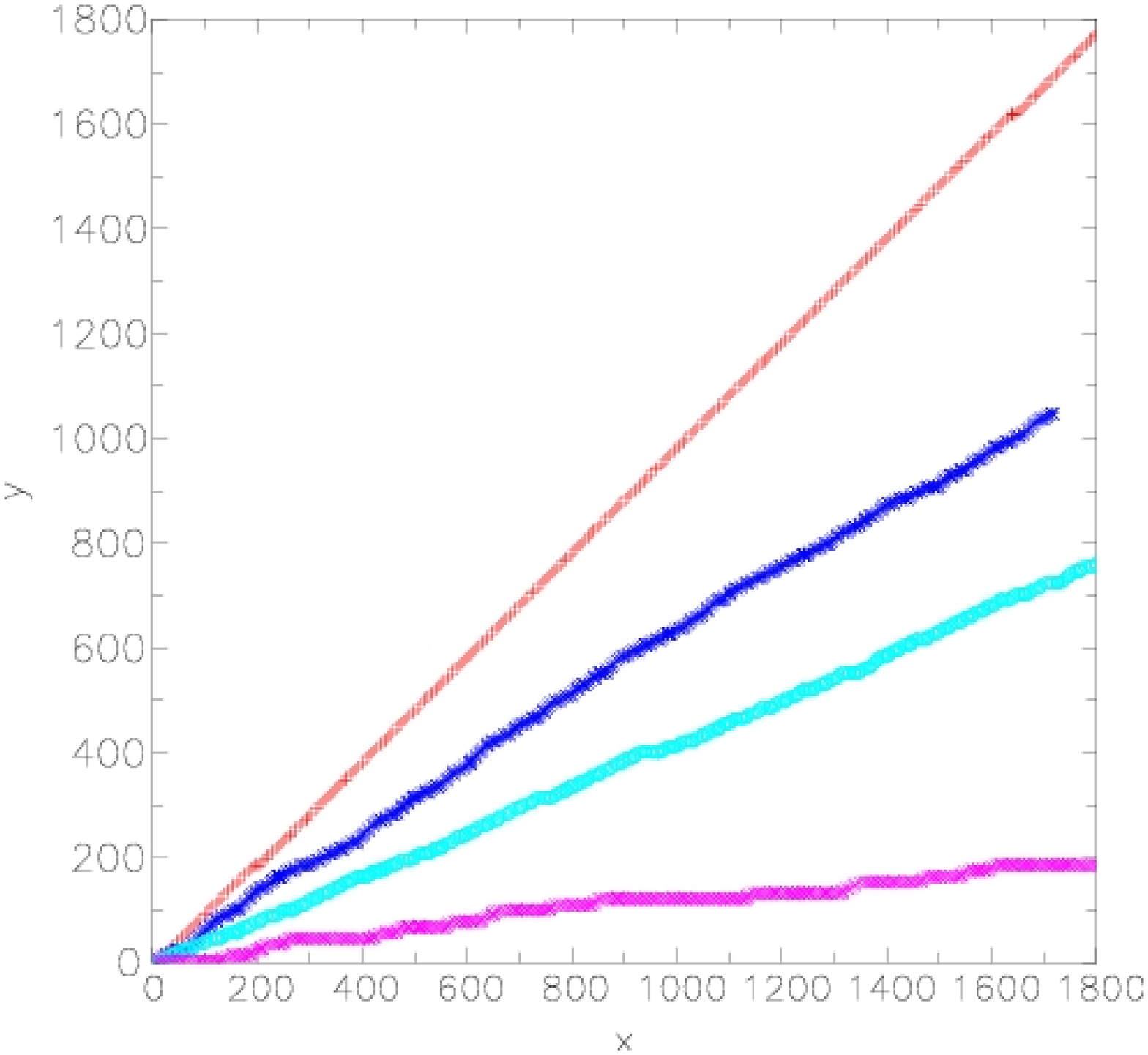}
\end{center}
\caption{\label{fig:coarsetraj} (Color online)
Initial and global trajectories of a particle forced at 
$F_0=100\epsilon/\sigma$ for angles $\theta=36$ (red), 30 (blue), 26
(cyan) and 22$^\circ$ (magenta), top to bottom.}
\end{figure}

The effects of molecular diffusion are
illustrated in Fig.~\ref{fig:finetraj} which shows a fine-scale view of the
initial particle motion at low and high $F_0$.  As expected, the motion between obstacles 
tends toward a straight line, and the motion around obstacles follows their
shape more closely at larger $F_0$;  note that numerically $\rm{Pe}=2F_0$
here.  At larger scales, the shape of the trajectories is quite sensitive 
to locking phenomena. In Fig.~\ref{fig:coarsetraj} we display sample
trajectories at $F_0=100\epsilon/\sigma$ for forcing angles $\theta=36$, 30,
26 and 22$^\circ$, separately for intermediate and long time intervals of
about 500 and 5000$\tau$, respectively.  
When the force direction is close to a locking angle of
45$^\circ$ (not shown), the particle follows a straight path with detours around the 
obstacles, and when the forcing angle $\theta$ is near zero (not shown) the particle
bounces back and forth between two rows of obstacles.  These two limiting
behaviors persist for angles somewhat larger than 0$^\circ$ and somewhat
smaller than 45$^\circ$, respectively.  At intermediate angles in the middle
of this interval the motion is rather irregular:  segments of variable 
length along the locking directions at 45, 26.6 and 0$^\circ$.  These 
directions correspond to displacements which would geometrically intersect the
obstacle lattice with $\delta y:\delta x$ of 1:1, 1:2 and 1:$\infty$;  
the other lattice possibilities such as 1:3 or 2:3 do not seem to appear 
in the trajectories to any significant degree, even when the force is 
applied at just those values. 

\begin{figure}
\begin{center}
\includegraphics[width=0.45\linewidth]{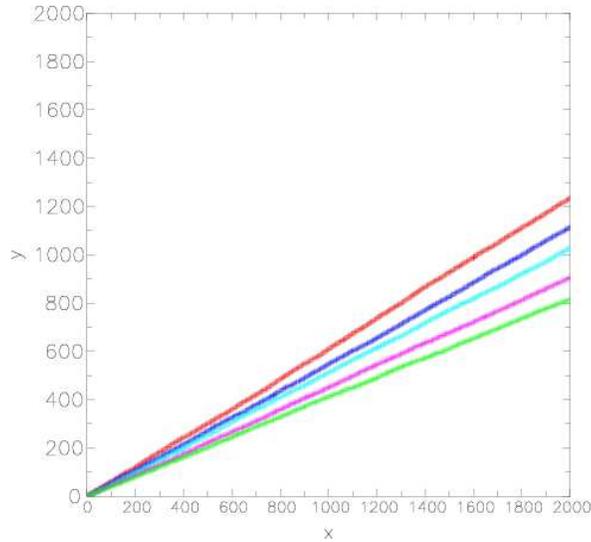}
\end{center}
\caption{\label{fig:3d} (Color online)
Trajectories of a sphere forced through a three-dimensional cubic array of
spherical obstacles, for forcing angles $\theta=(22.5,24.75,27,29.25,31.5)$
bottom to top (red,blue,cyan,magenta,green).  No locking occurs in this case.}
\end{figure}

At this point we note an important negative result:  locking in  
three-dimensional flows requires cylindrical obstacles which fully obstruct 
the motion in the third dimension, whereas a cubic array of localized 
obstacles has no systematic effect on the motion.  The reason is that the
deflection of the sphere around an obstacle after a close approach should be
independent of $z$ in order for the process to repeat coherently at
subsequent obstacles.  A cylindrical obstacle has this feature, but if the
obstacle is spherical the impact parameter depends on $z$ and will vary as
the particle diffuses in that direction. The direct evidence is given in 
Fig.~\ref{fig:3d} which shows the $x$-$y$ trajectories of a mobile sphere of 
radius 2$\sigma$ moving through a cubic array of spheres of the same radius  
spaced by 10.77$\sigma$. 
A force given by Eq.\ref{force} is applied with
$F_0=50\,\epsilon/\sigma$ at 
forcing angles $\theta=(22.5,24.75,27,29.25,31.5)$ and the inclination 
of the resulting trajectories is essentially the same: 
$\alpha=(22.27,24.51,27.08,29.19,31.52)$.
In this case a single realization at each angle suffices. 

\subsection{Locking behavior}

The locking behavior of the particle motion is determined by the relation
between the forcing angle $\theta$ and the trajectory angle $\alpha$,
defined as the angle between the end-to-end vector and the $x$-axis.
This angle could depend on the length of the trajectory, but in these
simulations we observe little sensitivity once the particle covers distances 
beyond a few hundred $\sigma$.  The results presented here are based on
simulations over a 5000$\tau$ time interval for flows with $\rm{Pe}>100$, which
cover at least this distance.  For each case we
average over five realizations, obtained by applying a force in the
(neutral) $z$-direction for different short intervals, to produce a new
initial position for the particle.  The resulting standard deviation in the
mean trajectory angle is one degree or less, and similarly the relative 
fluctuations in the other quantities discussed below is small.
We begin with simulations in the system
discussed above, using forces of magnitude $F_0=50$, 100 or 
200$\epsilon/\sigma$ oriented at angles 14 through 42$^\circ$.  As an
alternative to applying a fixed force to the particle which pulls it through
the solvent, in a separate series of simulations we apply a pressure gradient 
to the fluid which drives a flow and advects the particle.  In the latter
case, a force $\bf{g}$  is applied to each fluid atom at the same set of 
directions, with magnitude $g=0.1$, 0.25 or 0.5$\sigma/\tau^2$.  

\begin{figure}
\begin{center}
\includegraphics[width=0.45\linewidth]{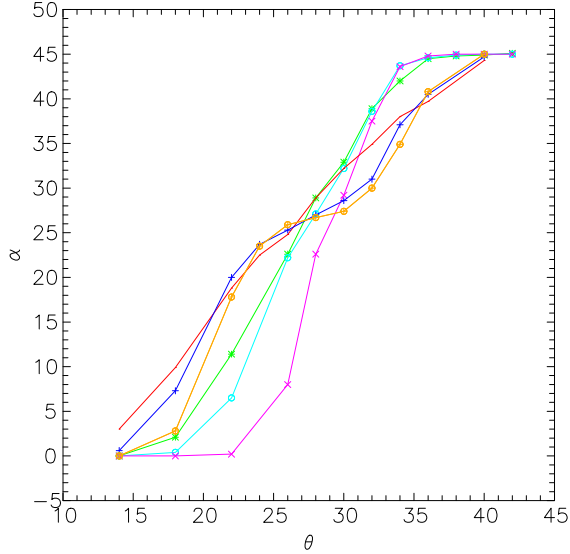}
\end{center}
\caption{\label{fig:angle} (Color online)
Trajectories angles $\alpha$ {\em vs}. forcing angle $\theta$ when the
particle is driven either by a constant force applied to the particle of
magnitude $F_0=50$ (green,*), 100 (cyan,o) or 200$\epsilon/\sigma$ (magenta,x), 
or else by an acceleration $g=0.1$ (red,$\cdot$), 0.25 (blue,+) or
0.5$\sigma/\tau^2$ (orange,$\rm{o}\!\!\!\!+$) applied to the solvent atoms.}
\end{figure}

The resulting variation
of the trajectory angle is shown in Fig.~\ref{fig:angle}, where obvious
locking behavior is seen:  all of the curves depart significantly from a
straight line at 45$^\circ$.  For smaller values of $\theta$ the trajectory
angle is approximately zero in all cases.  As expected, the locking 
behavior is more
pronounced at higher forcing values, corresponding to higher $\rm{Pe}$.  A more
surprising result is that a locking plateau -- a range of forcing angles
producing the same trajectory angle -- appears only at a subset of the
possibilities which occur in the ballistic and high $\rm{Pe}$ limits,\cite{HerrmannDrazer09,FrechetteDrazer09} 
at zero, 45 and possibly  26.6$^\circ$. The latter
near-plateau is most evident when the solvent is forced, at the two higher $g$
values.  Presumably the
difference arises from molecular diffusion, even for P\'eclet numbers of
several hundred which appear at the highest forcing values here. 

When higher values of force are applied to the particle a complication
arises:  the particle can become jammed into an obstacle and remain there 
for a variable length of time.  The reason for jamming is that both particle
and obstacle are
rigid solid lattices with (atomic) corners, and if a corner of the mobile 
particle enters the gap between two layers of the obstacle and the driving
force exceeds the typical LJ short-distance repulsive force, an equilibrium
may be established.  A simple (physically realistic) remedy is to introduce 
thermal motion for the atoms in the two solids, which alters the force 
balance and mobilizes the particle.  This technique was used in a few
test situations at $F_0=200$, with no significant change in the results, and
could be used to reach higher $\rm{Pe}$ values, but we have not pursued this
extension systematically.

\begin{figure}
\begin{center}
\includegraphics[width=0.45\linewidth]{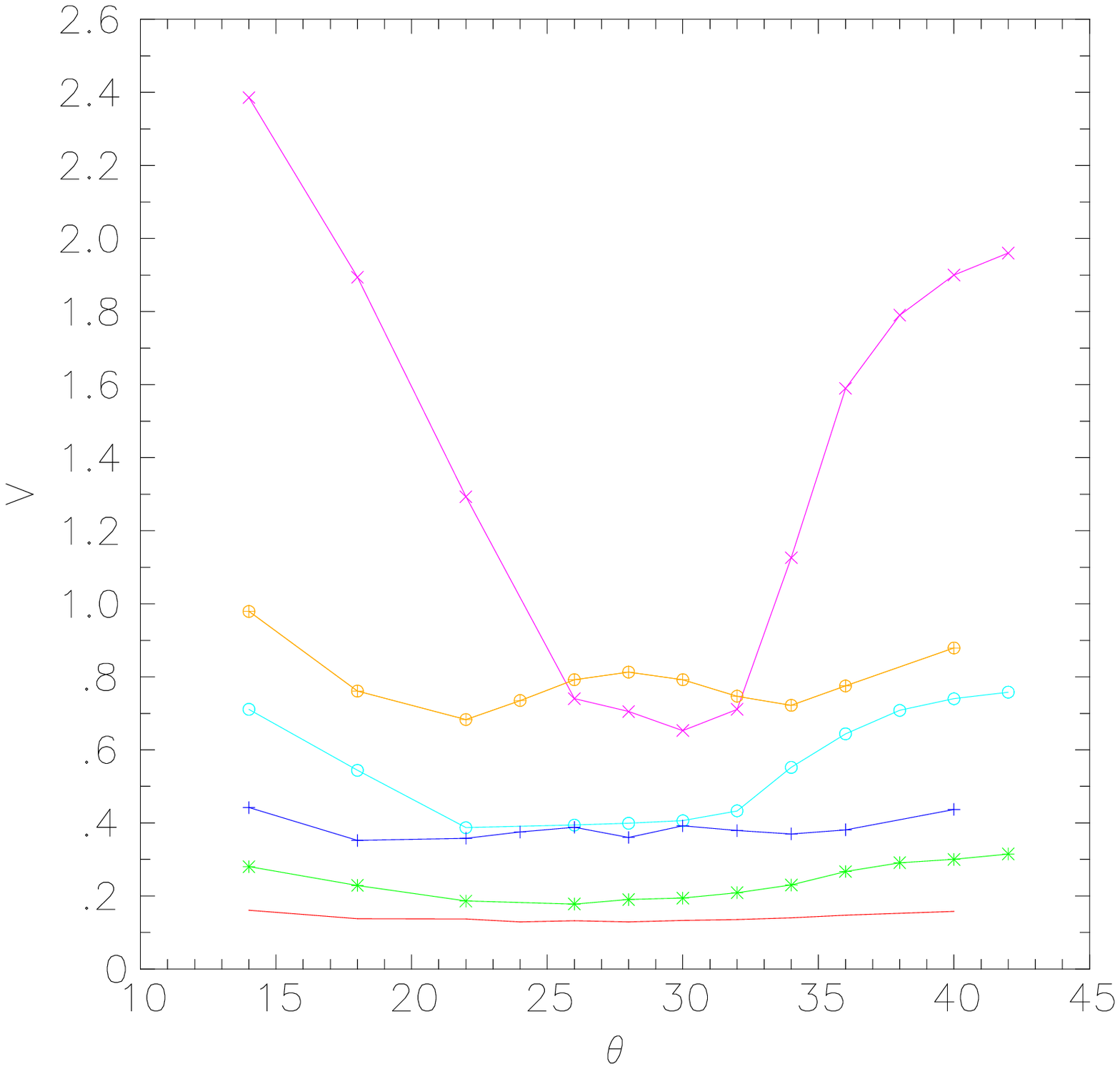}
\hspace{0.2in}
\includegraphics[width=0.45\linewidth]{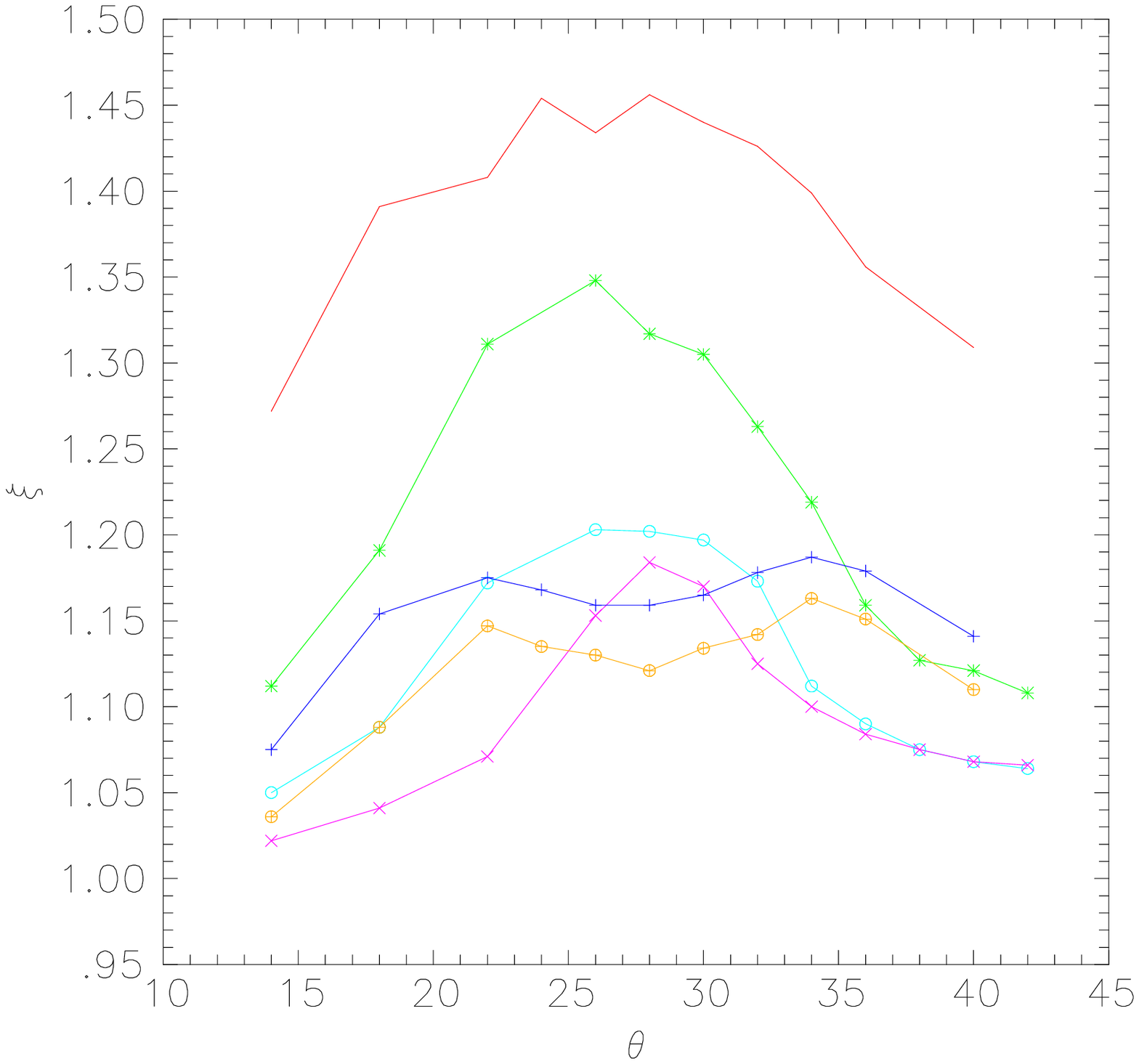}
\end{center}
\caption{\label{fig:v_xi} (Color online)
Average velocity (left) and excess length (right) for the trajectories in
Fig.~\ref{fig:angle}, plotted with the same colors and symbols.}
\end{figure}

In addition to the angle we measure the average velocity $V$ in each
trajectory, defined as the end-to-end distance in the $x$-$y$ plane 
divided by the elapsed time, 
and an ``excess length'' $\xi$, defined as the (approximate) path length
divided by the end-to-end distance.  The path length is approximated by
dividing the trajectory into straight segment of duration 1$\tau$ and
summing the (two-dimensional) segment displacements, and the significance of
$\xi$ is that it measures the deviation of a trajectory from a straight line
and thereby the degree to which an individual trajectory is in fact locked
at the angle $\alpha$.  The value of $\xi$ of course depends on the time
interval of the segments used to measure it and is not unique, but rather is
meant to provide a simple measure of ``dithering'' -- deviations from a
straight line along a trajectory.
The result for the average velocity and excess length are presented in
Fig.~\ref{fig:v_xi}
The velocities are largest and the excess length smallest when the forcing 
direction is along the $x$-axis and the particle simply moves through one
row of the lattice, and similar behavior is found when the forcing is close
to the 45$^\circ$ locking direction.  In the forced-particle cases, the 
velocity is smallest and the dithering is largest at intermediate forcing 
angles where the particle trajectory does not follow a single locking angle.
The forced-solvent simulations at $g=0.25$ and 0.5$\sigma/\tau^2$ show some
evidence for a plateau at the 26.6$^\circ$ locking angle, and correspondingly  
have local maxima in $V$ and local minima in $\xi$.

\begin{figure}
\begin{center}
\includegraphics[width=0.45\linewidth]{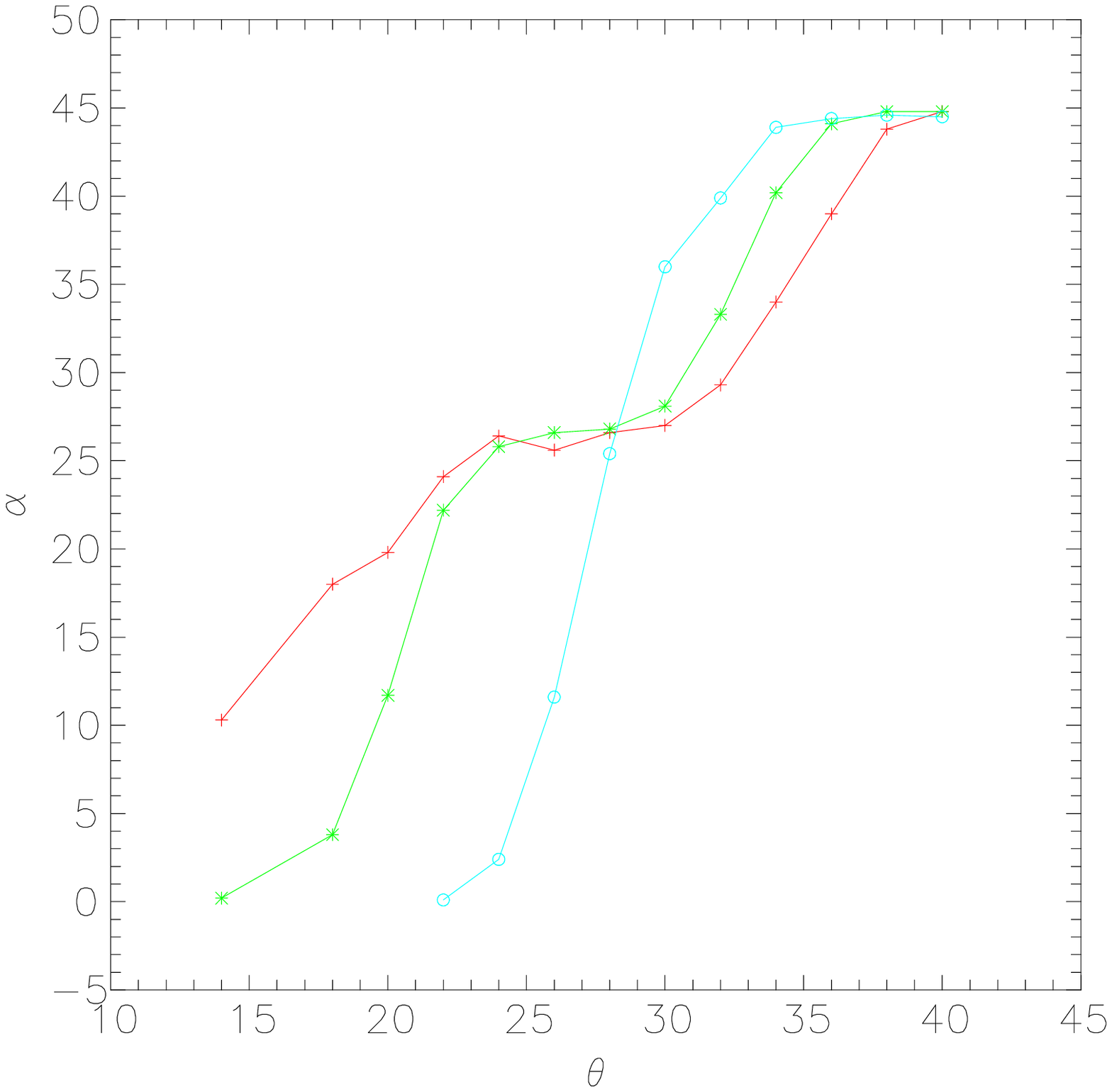}\\
\includegraphics[width=0.45\linewidth]{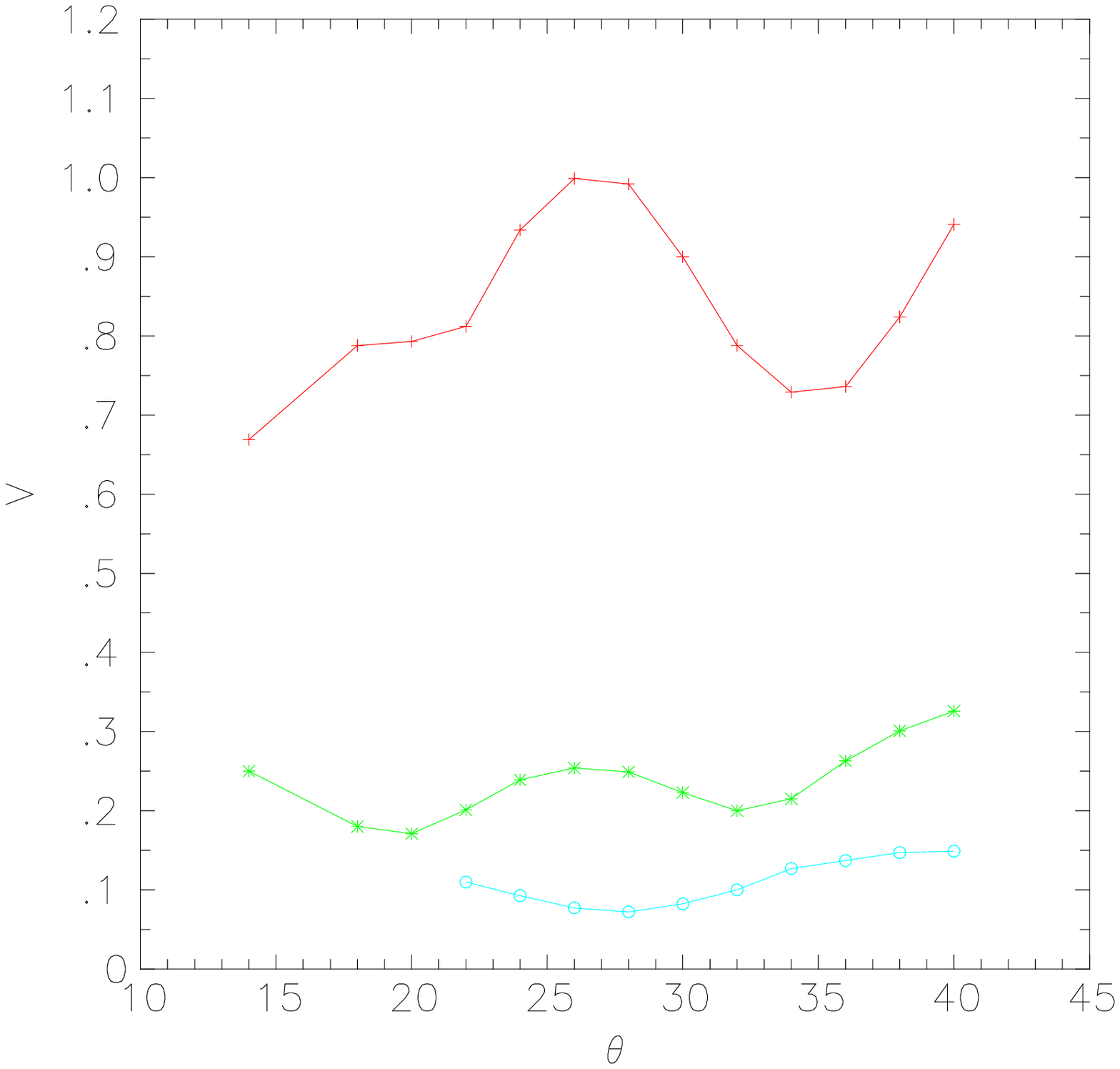}
\hspace{0.2in}
\includegraphics[width=0.45\linewidth]{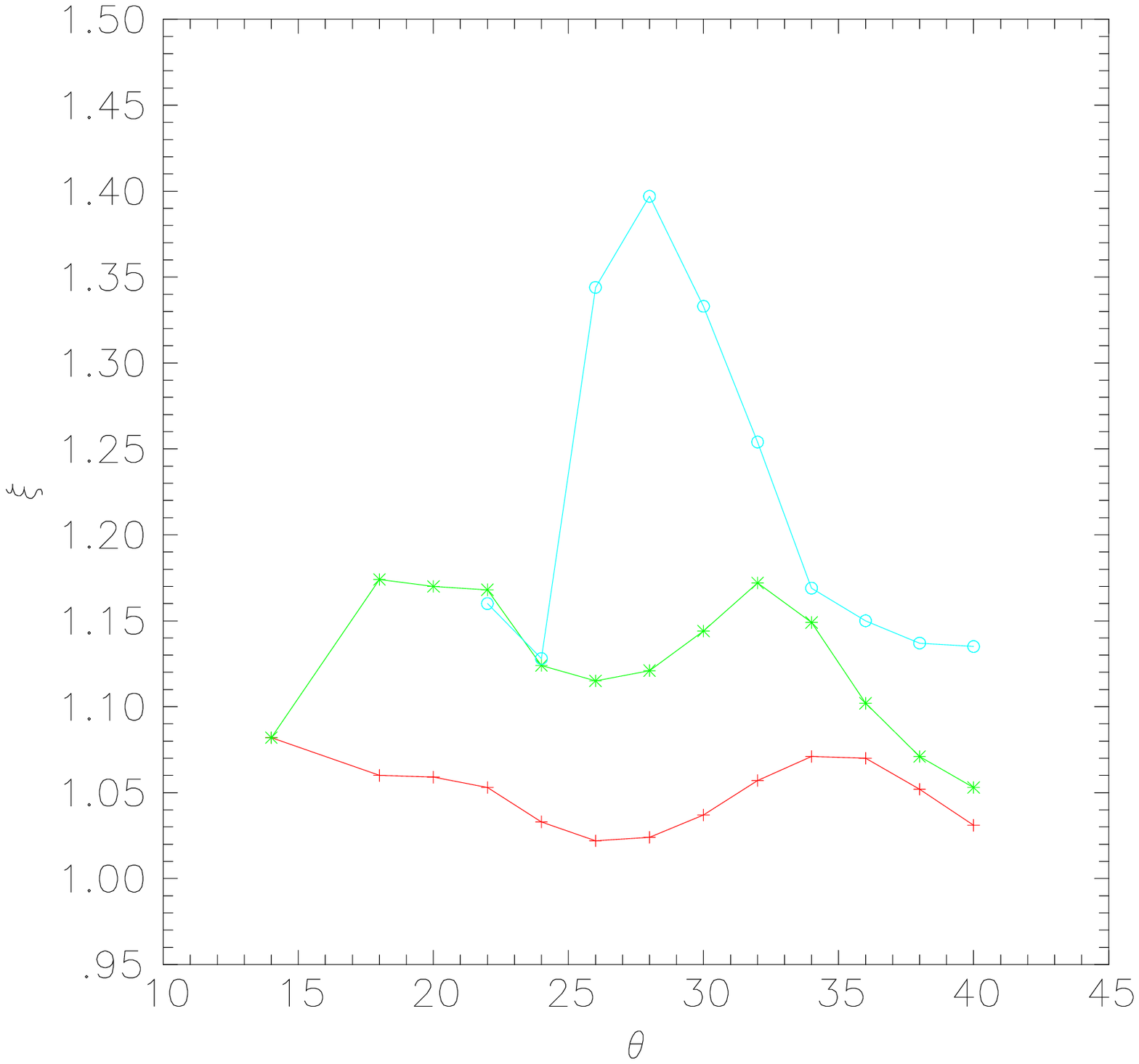}
\end{center}
\caption{\label{fig:size} (Color online) Effects of size.
Trajectory angle (top), average velocity (left) and excess length (right) 
as a function of forcing angle for particles of radius 2 (red,+), 4
(green,*) and 5.5 (cyan,o) in the larger-spacing obstacle lattice.}
\end{figure}

Next, we consider the effect of varying the geometry.  We compare the
behavior of particle of radius 2, 4 and 5.5$\sigma$, in a lattice
of cylinders of the previous radius 2 but a larger spacing 17.2$\sigma$ 
in order to accommodate the larger particles.  A force $F_0=100$ is applied
to the particle in each case.  In Fig.~\ref{fig:size}
we show the trajectory angle, average velocity and excess length 
as a function of forcing angle.
In this situation, we again see that the trajectories do not simply follow
the force direction, and at least for the two smaller particles the presence
of three locking plateaus.  In comparison to the previous case, the larger
obstacle spacing leads to a 26.6$^\circ$ plateau for particles of radius 2,
which suggest that a looser obstacle lattice, or more precisely a larger
ratio of obstacle spacing to particle size, may improve locking. 
The overall velocity decreases with particle size, reflecting the increase
in the hydrodynamic drag coefficient with radius, and the variation with
angle shows similar trends as in the previous geometry, high velocities
and low $\xi$ 
near the locking direction at 45$^\circ$  and near 26.6$^\circ$ for the two
smaller particle sizes where a locking plateau is present.  The radius 4 
simulations, which extend into the 0$^\circ$ locking regime, show a velocity
maximum and an excess length minimum there, consistent with the previous
system, but the other particle cases have not covered this regime.

\begin{figure}
\begin{center}
\includegraphics*[width=0.45\linewidth]{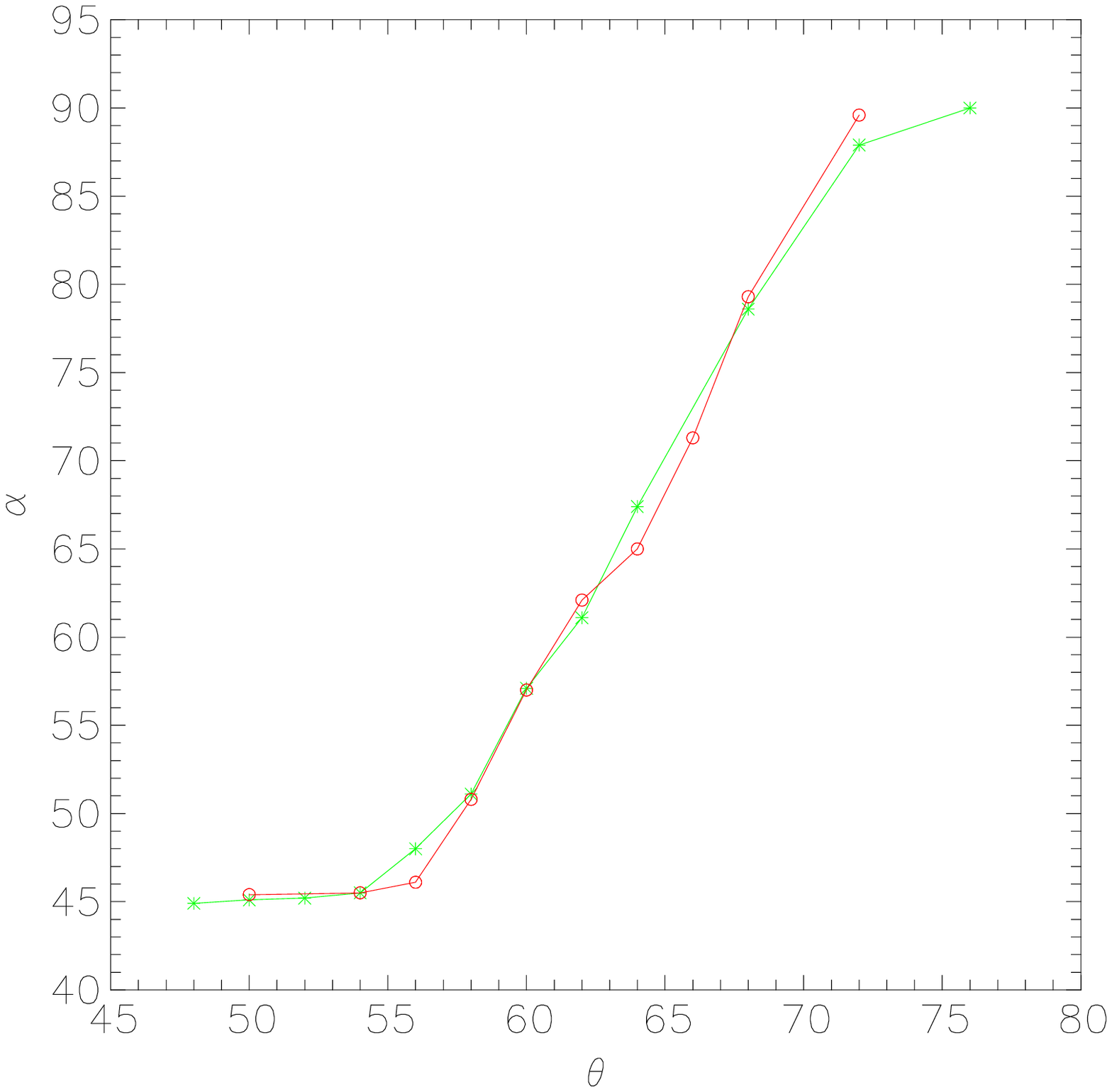}\\
\includegraphics*[width=0.45\linewidth]{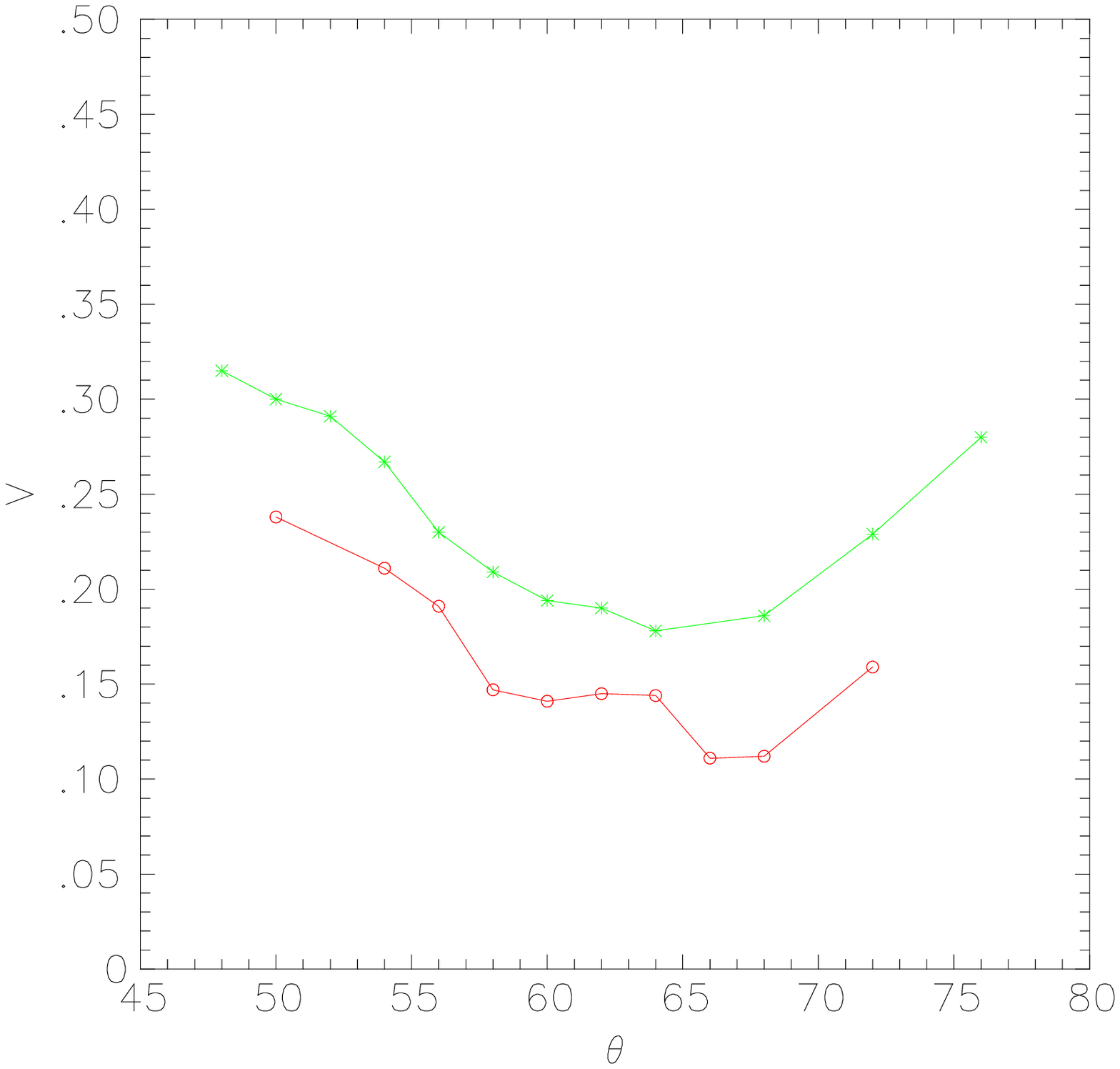}
\hspace{0.2in}
\includegraphics*[width=0.45\linewidth]{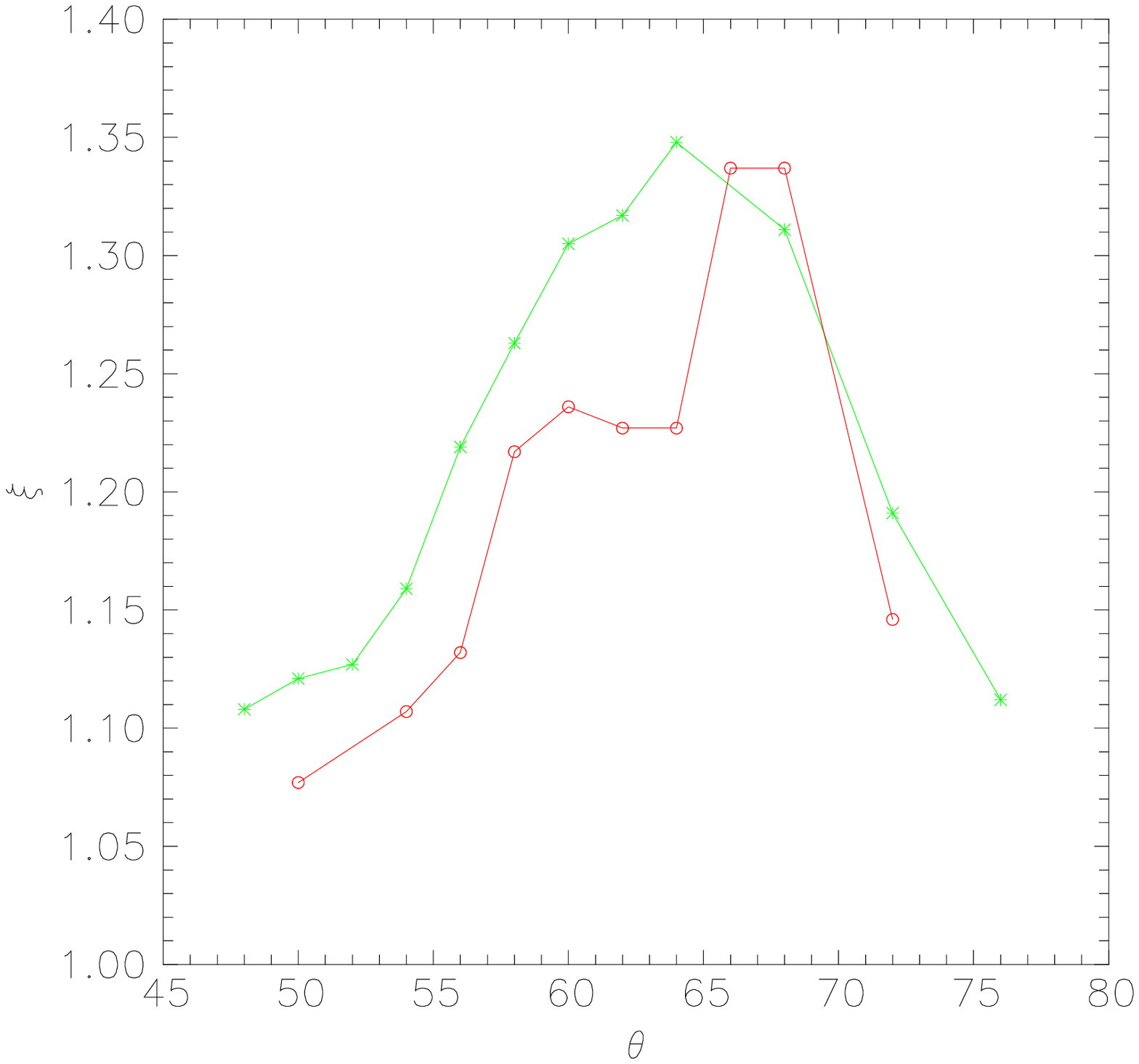}
\end{center}
\caption{\label{fig:double} (Color online) Effects of roughness.
Trajectory angle (top), average velocity (left) and excess length (right)
as a function of forcing angle for particles and obstacles of radius 2
spaced by 10.77 (green,*) at force $F_0=50$, compared to a system doubled 
in all sizes at force $F_0=100$ (red,o).}
\end{figure}

We have also varied the particle and obstacle size so as to explore the
effects of surface roughness.  The scale of the roughness is controlled by the
size of the atoms comprising the particle and obstacles, so if we double
their radii we have reduced the {\em relative} roughness by a half.  On
the other hand, if at the same time we double the obstacle spacing 
we nominally have scaled the system up in size by a factor of two, and to
the extent that these simulations are in the Stokes flow regime (the
suspending fluid's viscosity is about 2$m/(\sigma\tau)$ so in these 
simulations the Reynolds number is O(1)), the hydrodynamics is unchanged.
In Fig.~\ref{fig:double} we compared the results of doubling all sizes and
applying a force of magnitude 100 to the original radius 2 case at a force
of 50;  doubling the particle size approximately doubles the hydrodynamic 
drag so as to provide a fair comparison.  As seen in the figure, there is  
no significant change in the trajectory angle, and approximate agreement in 
the velocities and excess length.  Theses results suggest that it is the
absolute scale rather than the relative amount of roughness which controls
the behavior.

\subsection{Deterministic behavior in the Stokes limit}
In previous work we investigated the high $\rm{Pe}$ limit by means of Stokesian 
dynamics (SD) simulations, and showed the presence of directional locking.\cite{FrechetteDrazer09}
Stokesian dynamics simulations incorporate hydrodynamic interactions between 
the suspended particles and the solid obstacles in the limit of zero  
Reynolds number.
In addition, a short-range repulsive force is typically introduced in 
SD simulations to qualitatively model the effect of non-hydrodynamic interactions, 
such as irreversible forces originating from surface roughness. In fact,
the observed locking behavior was shown to depend on a small
parameter $\epsilon$ that determines the range of the repulsive force
or the relative magnitude of the surface roughness.

\begin{figure}
\centering
\includegraphics*[width=14cm]{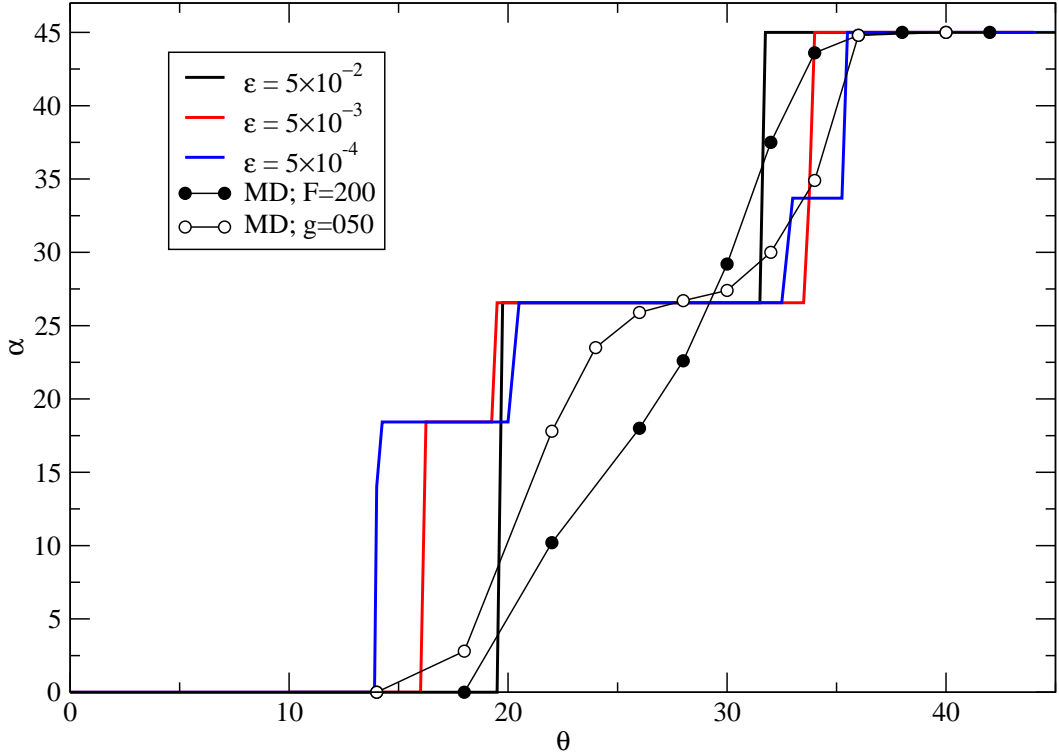}
\caption{\label{fig:devil} (Color online) Trajectory angles $\alpha$ 
as a function of the forcing direction $\theta$. The solid lines correspond 
to Stokesian dynamics
simulations with roughness parameters as indicated. The symbols correspond 
to the  
molecular simulations in Fig. \ref{fig:angle} for  $F_0=200 \epsilon/\sigma$ 
(solid) and $g=0.5 \sigma/\tau^2$ (open).}
\end{figure}

Here, we compare SD simulations with the behavior of nanoparticles
at relatively large P\'eclet numbers. The obstacle lattice in the SD simulations 
is represented by an array of fixed spheres and the suspended
particle moves in the plane of the array (note that in the case of 
deterministic trajectories
we do not have the problem discussed in section \ref{ss:local} regarding 
the absence of
locking in the case of spherical obstacles given that the particle does 
not diffuse in the $z$ direction). 
The particles forming the obstacle lattice as well as the moving sphere 
have the same radius $a=1$ and the lattice spacing is $5a$,
analogous to the system used in the nanoparticles simulations.

\begin{figure}
\includegraphics*[width=14cm]{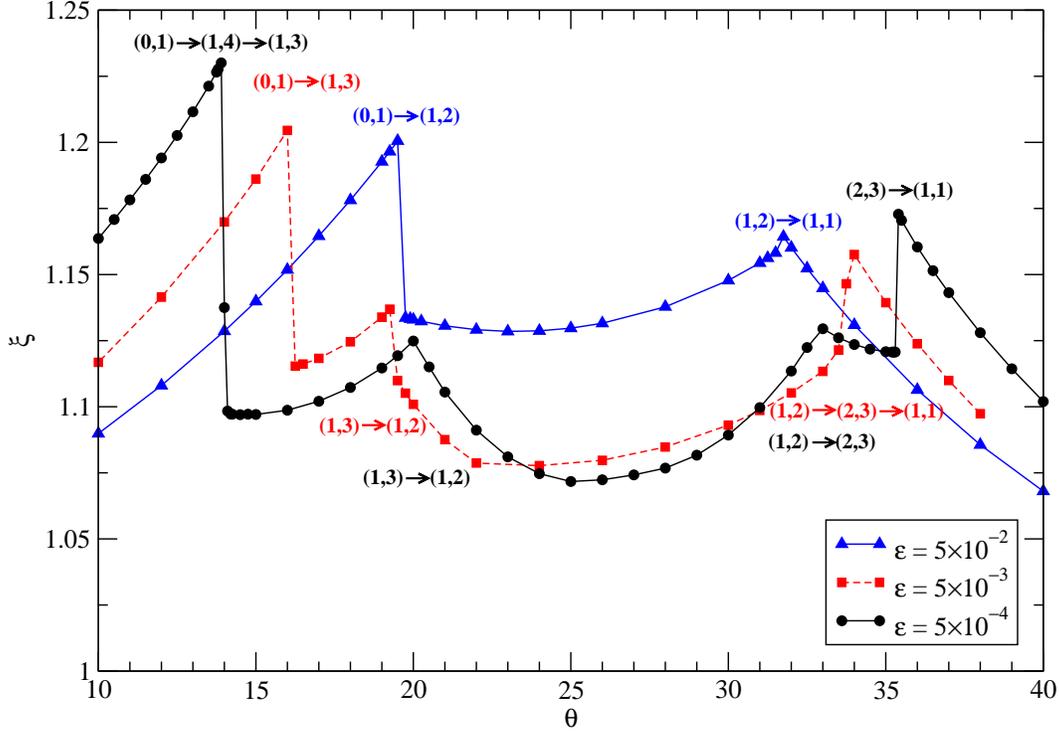}
\caption{\label{fig:length} (Color online) Excess length $\xi$ as a 
function of the forcing angle $\theta$ calculated
from Stokesian dynamics simulations for different values of the roughness 
parameter $\epsilon$ as indicated. 
The transitions between locking directions are indicated in the figure, 
where $(p,q)$ corresponds to motion
at an average angle $\alpha=\arctan(q/p)$.}
\end{figure}

The locking behavior is presented in Fig. \ref{fig:devil} for
different values of the roughness parameter, ranging from
$\epsilon=5\times10^{-4}$ to $\epsilon=5\times10^{-2}$. 
Typically, particles with larger roughness exhibit fewer locking angles, 
as seen in the figure.
We also compare the SD results with the
trajectory angles obtained for nanoparticles shown in Fig. \ref{fig:angle}
for the largest values of the P\'eclet numbers. We observe that the
migration angle of the nanoparticles is closer to the deterministic
trajectories corresponding to the largest value of the roughness
parameter.  It is also clear that the gravity driven nanoparticles
exhibit a plateau similar to that observed in the SD simulations
for intermediate forcing angles. On the other hand, the force-driven
nanoparticles seem to transition directly from $\alpha=0^\circ$ to
$\alpha=45^\circ$.  This behavior corresponds to deterministic
trajectories with values of the roughness parameter larger than
those considered here.

\begin{figure}
\includegraphics*[width=14cm]{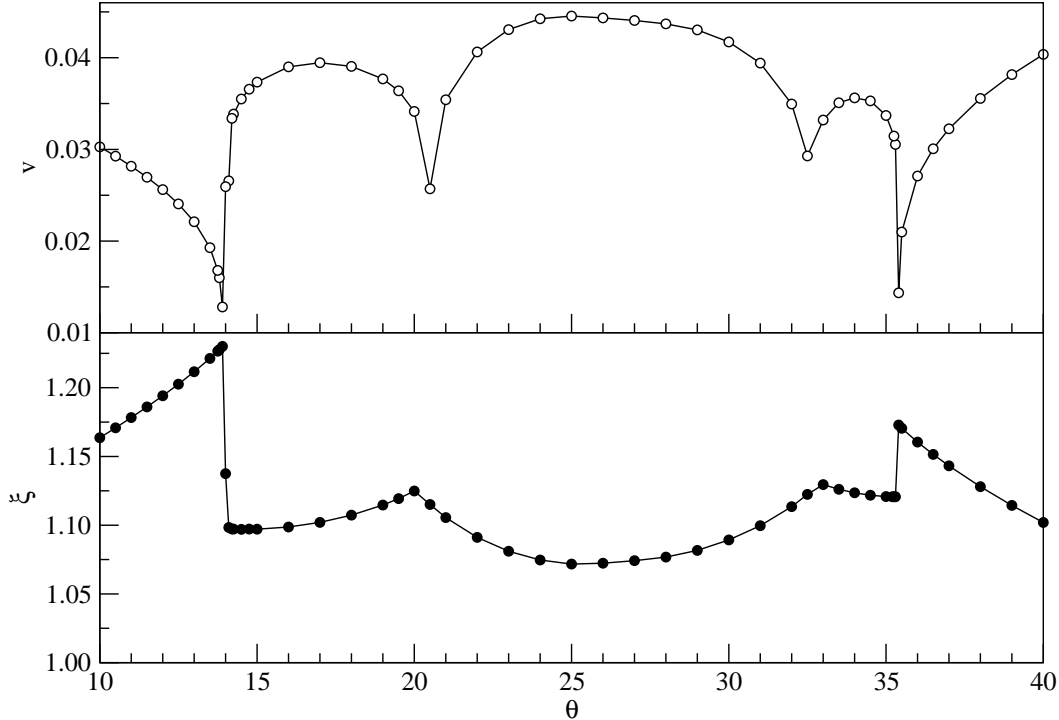}
\caption{\label{fig:lengthvel} Average velocity (top) and excess 
length (bottom) for the trajectories 
shown in Fig. \ref{fig:length} for the smallest value of the 
roughness parameter, $\epsilon=5\times10^{-4}$.}
\end{figure}

The variation of the excess length measured in the SD simulations is
also analogous to the one observed in the case of nanoparticles, 
but exhibits sharper transitions between locking angles. In Fig. \ref{fig:length} we
present the excess length as a function of the driving direction
for the different values of the roughness parameter considered in
Fig. \ref{fig:devil}. Similar to what we observe in the transport
of nanoparticles at large Pe, the excess length is smaller when the
particles move either at $\alpha=0^\circ$ or $\alpha=45^\circ$. At
intermediate angles the excess length displays sharp jumps or local
maxima at the forcing angles corresponding to transitions between
two locking directions. Similar dependence is observed in the motion
of nanoparticles presented in Figs. \ref{fig:v_xi} and \ref{fig:size}.
In the cases in which there is a plateau at intermediate driving
angles the excess length displays two clear maxima.  On the other
hand, in the cases in which $\alpha=0^\circ$ and $45^\circ$ are the
only locking directions, there is a single maximum in the excess
length around the transition angle.

The deterministic nature of the SD simulations also allowed us to
investigate the observed correlation between the average velocity
and the excess length in the trajectory of the particles. In Fig.
\ref{fig:lengthvel} we see that, in fact, they are highly correlated,
with local maxima in the excess length corresponding to minima in
the average velocity of the suspended particles. In fact, the slowest
regions of a given trajectory occur when the particle moves around
a solid obstacle with small particle-obstacle separations. On the
other hand, the fastest portions of any trajectory are those in
which hydrodynamic interactions with the obstacles are small and
particles move close to a straight line. As a result, small values
of the excess length, which correspond to trajectories closer to a
straight line, also indicate faster motion of the particles, and
vice versa.

\subsection{Suspensions}

\begin{figure}
\begin{center}
\includegraphics[width=0.45\linewidth]{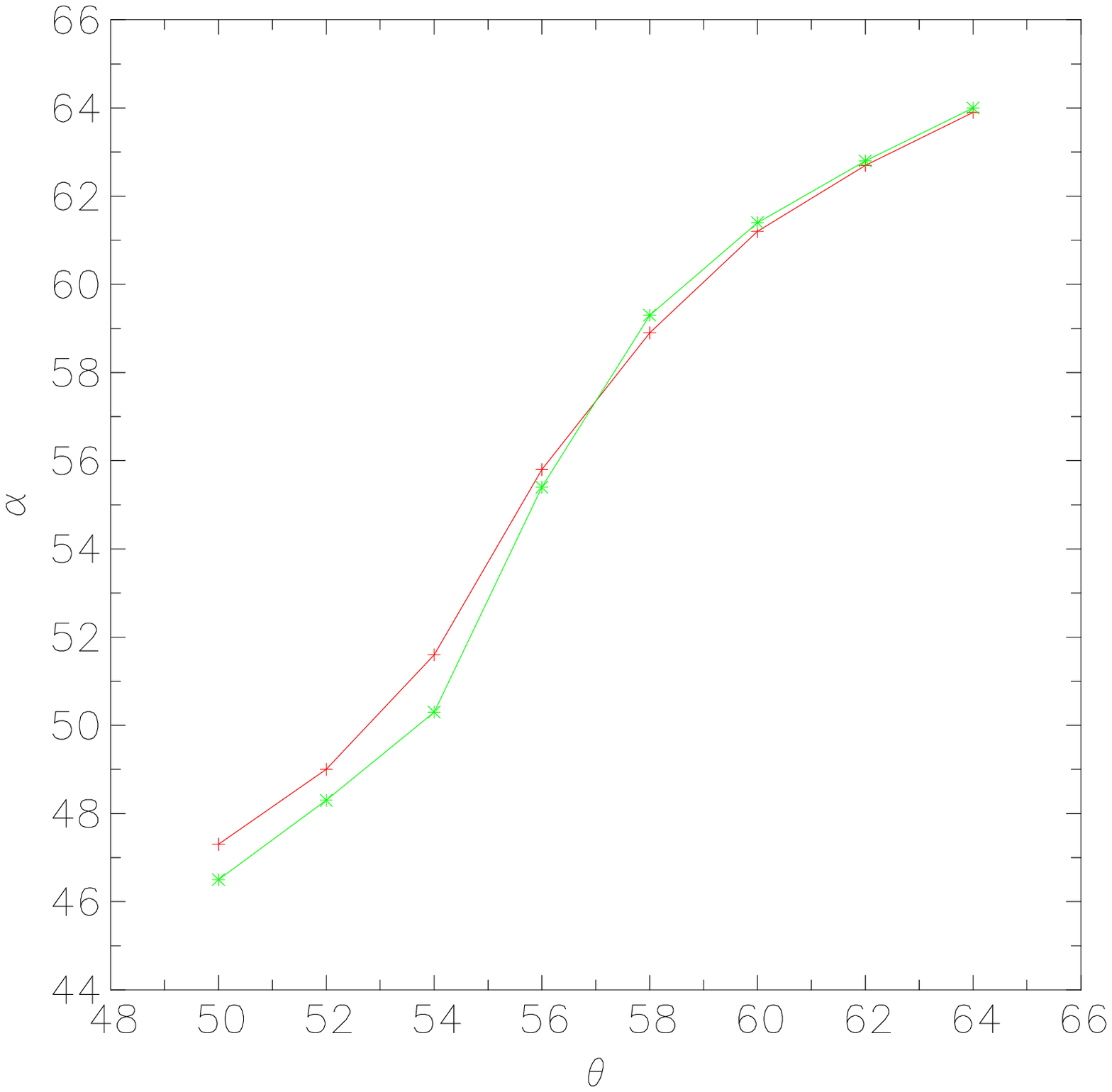}
\hspace{0.2in}
\includegraphics[width=0.45\linewidth]{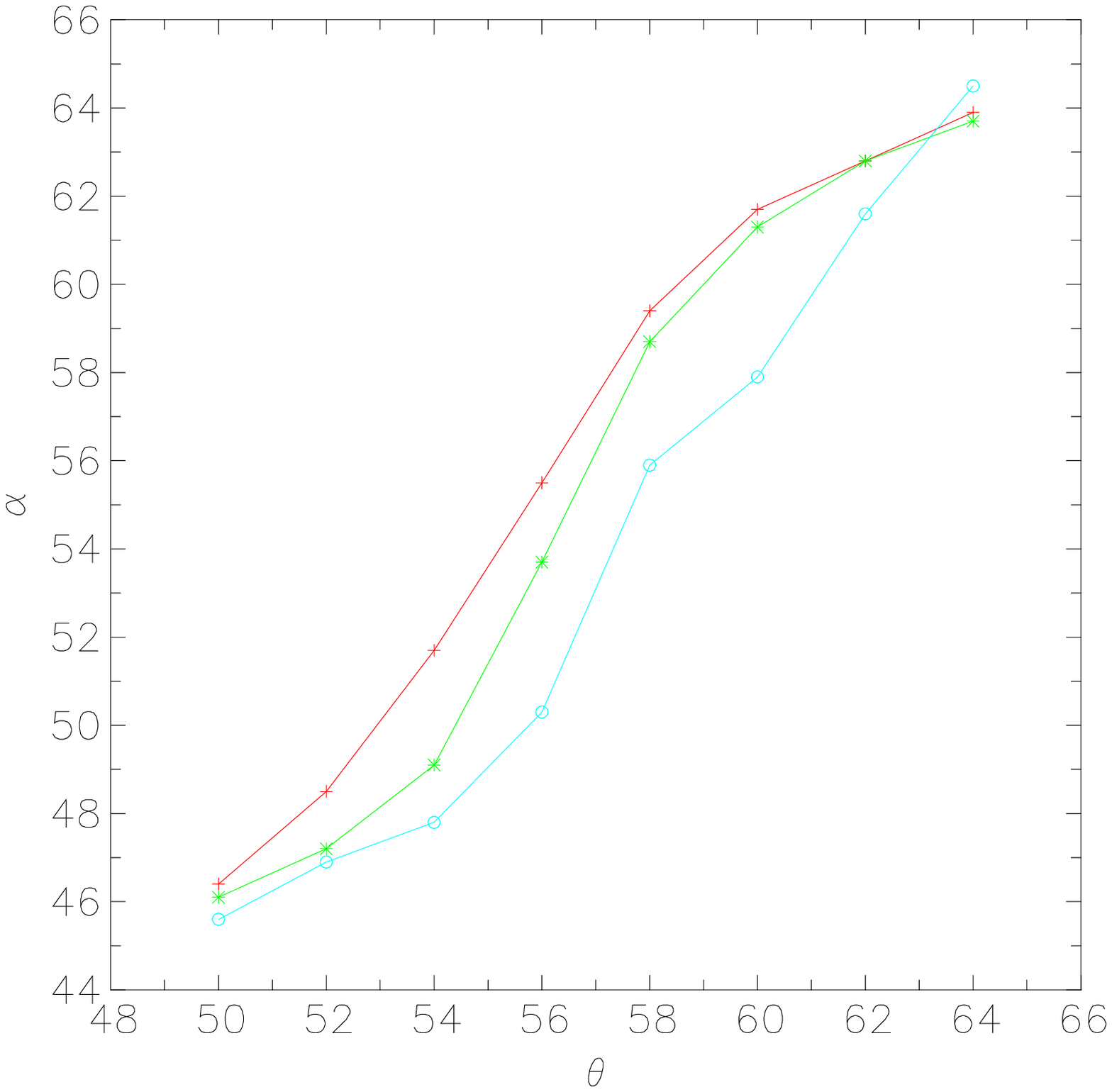}
\end{center}
\caption{\label{fig:susp} (Color online) Trajectory angle {\em vs}.
forcing angle for suspensions.  Left: ``small'' and right: ``big'' suspension, 
respectively;  with radius 2 (red,+0), radius 3 (green,*) and radius 4 
(cyan,o) particles displayed separately.}
\end{figure}

Given that the trajectories of a single isolated particle in an obstacle 
array are not collinear with an applied force and may exhibit locking
behavior, a natural question is whether this behavior persists in the
presence of other mobile particles.  The basic argument for locking
is based on the behavior of a single particle approaching a single obstacle 
but multiple particles in general introduce additional interactions which
may alter the result.  We have therefore repeated the above calculations for
{\em suspensions} of particles.  Two cases are considered, both using
obstacles of radius 2 spaced by 17.2, a ``small'' simulation involving a
3$\times$3 unit cell containing 20 mobile particles of radius 2 and 20 of
radius 3, and a ``large'' simulation with a 5$\times$5 unit cell, 30
particle of radius 2, 30 of radius 3 and 10 of radius 4.  The suspensions
are fairly dilute, with particle volume fractions of 6.6\% and 3.85\% for 
the small and big cases, respectively. In Fig.~\ref{fig:susp} we show the
trajectory angle for each size of particle in the two systems, when a force
of magnitude $F_0=100$ is applied.
Although the obstacles evidently have an influence on the trajectories, the
effect is considerably reduced in comparison to the simulations involving a 
single mobile particle.  The interactions between the particles alter the
motion around the obstacles sufficiently to prevent the accumulation of
identical individual displacements.  On the positive side, there is a clear
sensitivity to the particle radius in the ``big'' case, at lower total
concentration, which suggest that using flow through obstacles as a
separation technique is viable for sufficiently dilute suspensions. 

\section{Conclusions}

We have used molecular dynamics simulations to expand the range of previous 
calculations of locking behavior in flows past obstacle arrays.  The new
ingredients of the problem considered here are realistic surface roughness,
variable particle and obstacle sizes, and a comparison of applying a force
directly to the particle with forcing the solvent fluid.  

We first considered the behavior of trajectories as a function of P\'eclet
number, and observed that even at $\rm{Pe}=10$ the paths of driven particles
resembled those at high $\rm{Pe}$. At general forcing angles, trajectories
consisted of segments of variable length aligned along the commensurate 
locking directions of the obstacle lattice, but near the more robust locking 
directions (45$^\circ$ in particular) straighter sinusoidal paths
predominated.  We also verified the theoretical expectation that cylindrical   
obstacles are needed to produce locking behavior, by simulating flow through
a three-dimensional lattice of spherical obstacles.
We then considered the global particle motion for various forcing strengths
and orientations and for two types for forcing -- driving the particle
directly and driving the suspending fluid.  The degree of locking was found
to increase with $\rm{Pe}$ and to be different for the two forcing methods, but for
$\rm{Pe}=O(100)$ did not show the devil's staircase plateaus seen in the
$\rm{Pe}\to\infty$ limit.  These results were compared in detail to those of
Stokesian dynamics calculations.  Examination of the behavior of particles 
of different 
size in the same obstacle lattice suggests that smaller particles may lock
more effectively.  A follow-up study of the motion of multi-particle suspensions
through obstacles verified that the trajectory angles depend on particle size 
and implies that these may be used as a separation device.   Lastly, we
considered larger particles and obstacles made of more atoms, which leaves
the absolute scale of the roughness (roughly one atom) unchanged but reduces  
its scale relative to the particle size, and found no significant variation.

The general result is that independent of these variations, the 
simulations {\em all} show trajectory locking, in the sense that the
average orientation of the trajectory differs from the direction in which
the force is applied.  Clear effects are seen for all sizes and both
forcing mechanisms, once the P\'eclet number is O(10) or more.  However, the 
sharp step-wise transitions which occur in the absence of Brownian motion 
are smoothed by molecular diffusion and most of the locking angles 
observed in the convective limit, shown for example in Fig.~\ref{fig:devil}, 
are not apparent at finite P\'eclet number.     
These results are subject to the usual limitation of molecular dynamics
calculations:  small sizes and short times in comparison to many laboratory
studies and to continuum
methods which do not track atomic motions, but the simulations nonetheless
exhibit Navier-Stokes behavior and provide relevant results.  Furthermore,
for actual nanoparticle applications the simulations operate directly at the
relevant scales.

This material is partially based upon work supported by the National Science
Foundation under Grant No. CBET-0731032.

%\bibliography{mdlock}

\begin{thebibliography}{10}%
\makeatletter
\providecommand \@ifxundefined [1]{%
 \ifx #1\undefined \expandafter \@firstoftwo
 \else \expandafter \@secondoftwo
\fi
}%
\providecommand \@ifnum [1]{%
 \ifnum #1\expandafter \@firstoftwo
 \else \expandafter \@secondoftwo
\fi
}%
\providecommand \enquote [1]{``#1''}%
\providecommand \bibnamefont  [1]{#1}%
\providecommand \bibfnamefont [1]{#1}%
\providecommand \citenamefont [1]{#1}%
\providecommand\href[0]{\@sanitize\@href}%
\providecommand\@href[1]{\endgroup\@@startlink{#1}\endgroup\@@href}%
\providecommand\@@href[1]{#1\@@endlink}%
\providecommand \@sanitize [0]{\begingroup\catcode`\&12\catcode`\#12\relax}%
\@ifxundefined \pdfoutput {\@firstoftwo}{%
 \@ifnum{\z@=\pdfoutput}{\@firstoftwo}{\@secondoftwo}%
}{%
 \providecommand\@@startlink[1]{\leavevmode}%
 \providecommand\@@endlink[0]{}%
}{%
 \providecommand\@@startlink[1]{%
  \leavevmode
  \pdfstartlink
   attr{/Border[0 0 1 ]/H/I/C[0 1 1]}%
   user{/Subtype/Link/A<</Type/Action/S/URI/URI(#1)>>}%
  \relax
 }%
 \providecommand\@@endlink[0]{\pdfendlink}%
}%
\providecommand \url  [0]{\begingroup\@sanitize \@url }%
\providecommand \@url [1]{\endgroup\@href {#1}{\urlprefix}}%
\providecommand \urlprefix [0]{URL }%
\providecommand \Eprint[0]{\href }%
\@ifxundefined \urlstyle {%
  \providecommand \doi [1]{doi:\discretionary{}{}{}#1}%
}{%
  \providecommand \doi [0]{doi:\discretionary{}{}{}\begingroup
  \urlstyle{rm}\Url }%
}%
\providecommand \doibase [0]{http://dx.doi.org/}%
\providecommand \Doi[1]{\href{\doibase#1}}%
\providecommand \selectlanguage [0]{\@gobble}%
\providecommand \bibinfo [0]{\@secondoftwo}%
\providecommand \bibfield [0]{\@secondoftwo}%
\providecommand \translation [1]{[#1]}%
\providecommand \BibitemOpen[0]{}%
\providecommand \bibitemStop [0]{}%
\providecommand \bibitemNoStop [0]{.\EOS\space}%
\providecommand \EOS [0]{\spacefactor3000\relax}%
\providecommand \BibitemShut [1]{\csname bibitem#1\endcsname}%
%</preamble>
\bibitem{FrechetteDrazer09}%
  \BibitemOpen
  \bibfield{author}{%
  \bibinfo {author} {\bibfnamefont{J}~\bibnamefont{Frechette}}\ and\ \bibinfo
  {author} {\bibfnamefont{G}~\bibnamefont{Drazer}},\ }%
  \bibfield{title}{%
  \enquote{\bibinfo {title} {Directional locking and deterministic separation
  in periodic arrays},}\ }%
  \bibfield{journal}{%
  \bibinfo {journal} {J. Fluid Mech.}\ }%
  \textbf{\bibinfo {volume} {627}},\ \bibinfo {pages} {379} (\bibinfo {year}
  {2009})\BibitemShut{NoStop}%
\bibitem{BalvinFrechette09}%
  \BibitemOpen
  \bibfield{author}{%
  \bibinfo {author} {\bibfnamefont{M.}~\bibnamefont{Balvin}}, \bibinfo {author}
  {\bibfnamefont{E.}~\bibnamefont{Sohn}}, \bibinfo {author}
  {\bibfnamefont{T.}~\bibnamefont{Iracki}}, \bibinfo {author}
  {\bibfnamefont{G.}~\bibnamefont{Drazer}},\ and\ \bibinfo {author}
  {\bibfnamefont{J.}~\bibnamefont{Frechette}},\ }%
  \bibfield{title}{%
  \enquote{\bibinfo {title} {Directional locking and the role of irreversible
  interactions in deterministic hydrodynamics separations in microfluidic
  devices},}\ }%
  \bibfield{journal}{%
  \bibinfo {journal} {Phys. Rev. Lett.}\ }%
  \textbf{\bibinfo {volume} {103}},\ \bibinfo {pages} {078301} (\bibinfo {year}
  {2009})\BibitemShut{NoStop}%
\bibitem{HuangSturm04}%
  \BibitemOpen
  \bibfield{author}{%
  \bibinfo {author} {\bibfnamefont{Lotien~Richard}\ \bibnamefont{Huang}},
  \bibinfo {author} {\bibfnamefont{Edward~C}\ \bibnamefont{Cox}}, \bibinfo
  {author} {\bibfnamefont{Robert~H}\ \bibnamefont{Austin}},\ and\ \bibinfo
  {author} {\bibfnamefont{James~C}\ \bibnamefont{Sturm}},\ }%
  \bibfield{title}{%
  \enquote{\bibinfo {title} {Continuous particle separation through
  deterministic lateral displacement},}\ }%
  \bibfield{journal}{%
  \bibinfo {journal} {Science}\ }%
  \textbf{\bibinfo {volume} {304}},\ \bibinfo {pages} {987} (\bibinfo {year}
  {2004})\BibitemShut{NoStop}%
\bibitem{DavisAustin06}%
  \BibitemOpen
  \bibfield{author}{%
  \bibinfo {author} {\bibfnamefont{J.~A}\ \bibnamefont{Davis}}, \bibinfo
  {author} {\bibfnamefont{D.~W}\ \bibnamefont{Inglis}}, \bibinfo {author}
  {\bibfnamefont{K.~J}\ \bibnamefont{Morton}}, \bibinfo {author}
  {\bibfnamefont{D.~A}\ \bibnamefont{Lawrence}}, \bibinfo {author}
  {\bibfnamefont{L.~R}\ \bibnamefont{Huang}}, \bibinfo {author}
  {\bibfnamefont{S.~Y}\ \bibnamefont{Chou}}, \bibinfo {author}
  {\bibfnamefont{J.~C}\ \bibnamefont{Sturm}},\ and\ \bibinfo {author}
  {\bibfnamefont{R.~H}\ \bibnamefont{Austin}},\ }%
  \bibfield{title}{%
  \enquote{\bibinfo {title} {Deterministic hydrodynamics: Taking blood
  apart},}\ }%
  \bibfield{journal}{%
  \bibinfo {journal} {Proc. Natl. Acad. Sci.}\ }%
  \textbf{\bibinfo {volume} {103}},\ \bibinfo {pages} {14779} (\bibinfo {year}
  {2006})\BibitemShut{NoStop}%
\bibitem{MortonAustin08}%
  \BibitemOpen
  \bibfield{author}{%
  \bibinfo {author} {\bibfnamefont{K.~J}\ \bibnamefont{Morton}}, \bibinfo
  {author} {\bibfnamefont{K.}~\bibnamefont{Loutherback}}, \bibinfo {author}
  {\bibfnamefont{D.~W}\ \bibnamefont{Inglis}}, \bibinfo {author}
  {\bibfnamefont{O.~K}\ \bibnamefont{Tsui}}, \bibinfo {author}
  {\bibfnamefont{J.~C}\ \bibnamefont{Sturm}}, \bibinfo {author}
  {\bibfnamefont{S.~Y}\ \bibnamefont{Chou}},\ and\ \bibinfo {author}
  {\bibfnamefont{R.~H}\ \bibnamefont{Austin}},\ }%
  \bibfield{title}{%
  \enquote{\bibinfo {title} {Crossing microfluidic streamlines to lyse, label
  and wash cells},}\ }%
  \bibfield{journal}{%
  \bibinfo {journal} {Lab chip}\ }%
  \textbf{\bibinfo {volume} {8}},\ \bibinfo {pages} {1448} (\bibinfo {year}
  {2008})\BibitemShut{NoStop}%
\bibitem{GreenMurthy09}%
  \BibitemOpen
  \bibfield{author}{%
  \bibinfo {author} {\bibfnamefont{James~V.}\ \bibnamefont{Green}}, \bibinfo
  {author} {\bibfnamefont{Milica}\ \bibnamefont{Radisic}},\ and\ \bibinfo
  {author} {\bibfnamefont{Shashi~K.}\ \bibnamefont{Murthy}},\ }%
  \bibfield{title}{%
  \enquote{\bibinfo {title} {Deterministic lateral displacement as a means to
  enrich large cells for tissue engineering},}\ }%
  \bibfield{journal}{%
  \bibinfo {journal} {Anal. Chem.}\ }%
  \textbf{\bibinfo {volume} {81}},\ \bibinfo {pages} {9178} (\bibinfo {month}
  {Nov.}\ \bibinfo {year} {2009})\BibitemShut{NoStop}%
\bibitem{KordaGrier02}%
  \BibitemOpen
  \bibfield{author}{%
  \bibinfo {author} {\bibfnamefont{Pamela~T.}\ \bibnamefont{Korda}}, \bibinfo
  {author} {\bibfnamefont{Michael~B.}\ \bibnamefont{Taylor}},\ and\ \bibinfo
  {author} {\bibfnamefont{David~G.}\ \bibnamefont{Grier}},\ }%
  \bibfield{title}{%
  \enquote{\bibinfo {title} {Kinetically {Locked-In} colloidal transport in an
  array of optical tweezers},}\ }%
  \bibfield{journal}{%
  \bibinfo {journal} {Phys. Rev. Lett.}\ }%
  \textbf{\bibinfo {volume} {89}},\ \bibinfo {pages} {128301} (\bibinfo {year}
  {2002})\BibitemShut{NoStop}%
\bibitem{MacDonaldDholakia03}%
  \BibitemOpen
  \bibfield{author}{%
  \bibinfo {author} {\bibfnamefont{M.~P.}\ \bibnamefont{{MacDonald}}}, \bibinfo
  {author} {\bibfnamefont{G.~C.}\ \bibnamefont{Spalding}},\ and\ \bibinfo
  {author} {\bibfnamefont{K.}~\bibnamefont{Dholakia}},\ }%
  \bibfield{title}{%
  \enquote{\bibinfo {title} {Microfluidic sorting in an optical lattice},}\ }%
  \bibfield{journal}{%
  \bibinfo {journal} {Nature}\ }%
  \textbf{\bibinfo {volume} {426}},\ \bibinfo {pages} {421} (\bibinfo {month}
  {Nov.}\ \bibinfo {year} {2003})\BibitemShut{NoStop}%
\bibitem{DholakiaMacDonald09}%
  \BibitemOpen
  \bibfield{author}{%
  \bibinfo {author} {\bibfnamefont{K.}~\bibnamefont{Dholakia}}, \bibinfo
  {author} {\bibfnamefont{Woei~Ming}\ \bibnamefont{Lee}}, \bibinfo {author}
  {\bibfnamefont{L.}~\bibnamefont{Paterson}}, \bibinfo {author}
  {\bibfnamefont{{M.P.}}\ \bibnamefont{{MacDonald}}}, \bibinfo {author}
  {\bibfnamefont{R.}~\bibnamefont{{McDonald}}}, \bibinfo {author}
  {\bibfnamefont{I.}~\bibnamefont{Andreev}}, \bibinfo {author}
  {\bibfnamefont{P.}~\bibnamefont{Mthunzi}}, \bibinfo {author}
  {\bibfnamefont{{C.T.A.}}\ \bibnamefont{Brown}}, \bibinfo {author}
  {\bibfnamefont{{R.F.}}\ \bibnamefont{Marchington}},\ and\ \bibinfo {author}
  {\bibfnamefont{{A.C.}}\ \bibnamefont{Riches}},\ }%
  \bibfield{title}{%
  \enquote{\bibinfo {title} {Optical separation of cells on potential energy
  landscapes: Enhancement with dielectric tagging},}\ }%
  \bibfield{journal}{%
  \bibinfo {journal} {IEEE J. Sel. Topics Quantum Electron.}\ }%
  \textbf{\bibinfo {volume} {13}},\ \bibinfo {pages} {1646} (\bibinfo {year}
  {2007})\BibitemShut{NoStop}%
\bibitem{LadavacGrier04}%
  \BibitemOpen
  \bibfield{author}{%
  \bibinfo {author} {\bibfnamefont{K.}~\bibnamefont{Ladavac}}, \bibinfo
  {author} {\bibfnamefont{K.}~\bibnamefont{Kasza}},\ and\ \bibinfo {author}
  {\bibfnamefont{D.~G.}\ \bibnamefont{Grier}},\ }%
  \bibfield{title}{%
  \enquote{\bibinfo {title} {Sorting mesoscopic objects with periodic potential
  landscapes: Optical fractionation},}\ }%
  \bibfield{journal}{%
  \bibinfo {journal} {Phys. Rev. E}\ }%
  \textbf{\bibinfo {volume} {70}},\ \bibinfo {pages} {010901} (\bibinfo {year}
  {2004})\BibitemShut{NoStop}%
\bibitem{RoichmanGrier07}%
  \BibitemOpen
  \bibfield{author}{%
  \bibinfo {author} {\bibfnamefont{Y.}~\bibnamefont{Roichman}}, \bibinfo
  {author} {\bibfnamefont{V.}~\bibnamefont{Wong}},\ and\ \bibinfo {author}
  {\bibfnamefont{D.~G.}\ \bibnamefont{Grier}},\ }%
  \bibfield{title}{%
  \enquote{\bibinfo {title} {Colloidal transport through optical tweezer
  arrays},}\ }%
  \bibfield{journal}{%
  \bibinfo {journal} {Phys. Rev. E}\ }%
  \textbf{\bibinfo {volume} {75}},\ \bibinfo {pages} {011407} (\bibinfo {year}
  {2007})\BibitemShut{NoStop}%
\bibitem{HerrmannDrazer09}%
  \BibitemOpen
  \bibfield{author}{%
  \bibinfo {author} {\bibfnamefont{John}\ \bibnamefont{Herrmann}}, \bibinfo
  {author} {\bibfnamefont{Michael}\ \bibnamefont{Karweit}},\ and\ \bibinfo
  {author} {\bibfnamefont{German}\ \bibnamefont{Drazer}},\ }%
  \bibfield{title}{%
  \enquote{\bibinfo {title} {Separation of suspended particles in microfluidic
  systems by directional locking in periodic fields},}\ }%
  \bibfield{journal}{%
  \bibinfo {journal} {Phys. Rev. E}\ }%
  \textbf{\bibinfo {volume} {79}},\ \bibinfo {pages} {061404} (\bibinfo {year}
  {2009})\BibitemShut{NoStop}%
\bibitem{ReichhardtNori99}%
  \BibitemOpen
  \bibfield{author}{%
  \bibinfo {author} {\bibfnamefont{C.}~\bibnamefont{Reichhardt}}\ and\ \bibinfo
  {author} {\bibfnamefont{F.}~\bibnamefont{Nori}},\ }%
  \bibfield{title}{%
  \enquote{\bibinfo {title} {Phase locking, devil's staircases, farey trees,
  and arnold tongues in driven vortex lattices with periodic pinning},}\ }%
  \bibfield{journal}{%
  \bibinfo {journal} {Phys. Rev. Lett.}\ }%
  \textbf{\bibinfo {volume} {82}},\ \bibinfo {pages} {414} (\bibinfo {year}
  {1999})\BibitemShut{NoStop}%
\bibitem{PeltonGrier04}%
  \BibitemOpen
  \bibfield{author}{%
  \bibinfo {author} {\bibfnamefont{M.}~\bibnamefont{Pelton}}, \bibinfo {author}
  {\bibfnamefont{K.}~\bibnamefont{Ladavac}},\ and\ \bibinfo {author}
  {\bibfnamefont{D.~G.}\ \bibnamefont{Grier}},\ }%
  \bibfield{title}{%
  \enquote{\bibinfo {title} {Transport and fractionation in periodic
  potential-energy landscapes},}\ }%
  \bibfield{journal}{%
  \bibinfo {journal} {Phys. Rev. E}\ }%
  \textbf{\bibinfo {volume} {70}},\ \bibinfo {pages} {031108} (\bibinfo {year}
  {2004})\BibitemShut{NoStop}%
\bibitem{ReichhardtReichhardt04}%
  \BibitemOpen
  \bibfield{author}{%
  \bibinfo {author} {\bibfnamefont{C.}~\bibnamefont{Reichhardt}}\ and\ \bibinfo
  {author} {\bibfnamefont{C.~J.~O}\ \bibnamefont{Reichhardt}},\ }%
  \bibfield{title}{%
  \enquote{\bibinfo {title} {Directional locking effects and dynamics for
  particles driven through a colloidal lattice},}\ }%
  \bibfield{journal}{%
  \bibinfo {journal} {Phys. Rev. E}\ }%
  \textbf{\bibinfo {volume} {69}},\ \bibinfo {pages} {041405} (\bibinfo {year}
  {2004})\BibitemShut{NoStop}%
\bibitem{LacastaLindenberg05}%
  \BibitemOpen
  \bibfield{author}{%
  \bibinfo {author} {\bibfnamefont{A.~M.}\ \bibnamefont{Lacasta}}, \bibinfo
  {author} {\bibfnamefont{J.~M.}\ \bibnamefont{Sancho}}, \bibinfo {author}
  {\bibfnamefont{A.~H.}\ \bibnamefont{Romero}},\ and\ \bibinfo {author}
  {\bibfnamefont{K.}~\bibnamefont{Lindenberg}},\ }%
  \bibfield{title}{%
  \enquote{\bibinfo {title} {Sorting on periodic surfaces},}\ }%
  \bibfield{journal}{%
  \bibinfo {journal} {Phys. Rev. Lett.}\ }%
  \textbf{\bibinfo {volume} {94}},\ \bibinfo {pages} {160601} (\bibinfo {year}
  {2005})\BibitemShut{NoStop}%
\bibitem{LacastaLindenberg06}%
  \BibitemOpen
  \bibfield{author}{%
  \bibinfo {author} {\bibfnamefont{A.~M}\ \bibnamefont{Lacasta}}, \bibinfo
  {author} {\bibfnamefont{M.}~\bibnamefont{Khoury}}, \bibinfo {author}
  {\bibfnamefont{J.~M}\ \bibnamefont{Sancho}},\ and\ \bibinfo {author}
  {\bibfnamefont{K.}~\bibnamefont{Lindenberg}},\ }%
  \bibfield{title}{%
  \enquote{\bibinfo {title} {Sorting of mesoscopic particles driven through
  periodic potential landscapes},}\ }%
  \bibfield{journal}{%
  \bibinfo {journal} {Mod. Phys. Lett. B}\ }%
  \textbf{\bibinfo {volume} {20}},\ \bibinfo {pages} {1427} (\bibinfo {year}
  {2006})\BibitemShut{NoStop}%
\bibitem{AllenTildesley87}%
  \BibitemOpen
  \bibfield{author}{%
  \bibinfo {author} {\bibfnamefont{M.~P.}\ \bibnamefont{Allen}}\ and\ \bibinfo
  {author} {\bibfnamefont{D.~J.}\ \bibnamefont{Tildesley}},\ }%
  \emph{\bibinfo {title} {Computer Simulation of Liquids}}\ (\bibinfo
  {publisher} {Clarendon Press},\ \bibinfo {address} {Oxford},\ \bibinfo {year}
  {1987})\BibitemShut{NoStop}%
\bibitem{EvansMurad77}%
  \BibitemOpen
  \bibfield{author}{%
  \bibinfo {author} {\bibfnamefont{Denis~J.}\ \bibnamefont{Evans}}\ and\
  \bibinfo {author} {\bibfnamefont{Sohail}\ \bibnamefont{Murad}},\ }%
  \bibfield{title}{%
  \enquote{\bibinfo {title} {Singularity free algorithm for molecular dynamics
  simulation of rigid polyatomics},}\ }%
  \bibfield{journal}{%
  \bibinfo {journal} {Mol. Phys.}\ }%
  \textbf{\bibinfo {volume} {34}},\ \bibinfo {pages} {327} (\bibinfo {year}
  {1977})\BibitemShut{NoStop}%
\end{thebibliography}

%

\end{document}